\documentclass[useAMS,usenatbib,usegraphicx]{mn2e}

\usepackage{amsmath}
\usepackage{amsfonts}
\usepackage{amssymb}

\newcommand{\aap}{A\&A} 
\newcommand{\mnras}{MNRAS} 
\newcommand{\apj}{ApJ} 
\newcommand{\araa}{ARA\&A} 
\newcommand{\apjs}{ApJS}
\newcommand{\aj}{AJ}
\newcommand{\pasp}{PASP} 
\newcommand{\ssr}{Space Sci. Rev.}
\newcommand{\qjras}{QJRAS}

\title[The interstellar media of early-type galaxies]
    {Galaxy Zoo: dust and molecular gas in early-type galaxies with prominent dust lanes}

\author[Sugata Kaviraj et al.]{Sugata Kaviraj,$^{1,2}$\thanks{E-mail: s.kaviraj@imperial.ac.uk; skaviraj@astro.ox.ac.uk} Yuan-Sen Ting,$^{2,3}$ Martin Bureau,$^{2}$ Stanislav S. Shabala,$^{2,4}$
\newauthor R. Mark Crockett,$^{2}$ Joseph Silk,$^{2}$ Chris Lintott,$^{2}$ Arfon Smith,$^{2}$ William C. Keel,$^{5}$
\newauthor Karen L. Masters,$^{6}$ Kevin Schawinski$^{7,9}$ and Steven P. Bamford$^{8}$\\\\
$^{1}$Blackett Laboratory, Imperial College London, London SW7 2AZ\\
$^{2}$Department of Physics, University of Oxford, Keble Road, Oxford, OX1 3RH\\
$^{3}$Ecole Polytechnique, 91128 Palaiseau Cedex, France\\
$^{4}$School of Mathematics \& Physics, Private Bag 37, University of Tasmania, Hobart 7001, Australia\\
$^{5}$Department of Physics and Astronomy, 206 Gallalee Hall, 514 University Blvd., University of Alabama, Tuscaloosa, AL 35487-0234, USA\\
$^{6}$Institute for Cosmology and Gravitation, University of Portsmouth, Dennis Sciama Building, Burnaby Road, Portsmouth PO1 3FX\\
$^{7}$Department of Physics, Yale University, New Haven, CT 06511, USA\\
$^{8}$School of Physics and Astronomy, University of Nottingham, University Park, Nottingham NG7 2RD \\
$^{9}$Einstein Fellow\\\\
This publication has been made possible by the participation of more than 250\,000 volunteers in the Galaxy Zoo project. Their \\
contributions are individually acknowledged at http://www.galaxyzoo.org/Volunteers.aspx.} 

\date{Accepted 2012 March 19. Received 2012 February 25; in original form 2011 July 26}
\pagerange{\pageref{firstpage}--\pageref{lastpage}} \pubyear{2012}

\begin{document}

\maketitle


\begin{abstract}
We explore the properties of dust and associated molecular gas in 352 nearby ($0.01<z<0.07$) early-type galaxies (ETGs) with prominent dust lanes, drawn from the Sloan Digital Sky Survey (SDSS). Two-thirds of these `dusty ETGs' (D-ETGs) are morphologically disturbed, which suggests a merger origin, making these galaxies ideal test beds for studying the merger process at low redshift. The D-ETGs preferentially reside in lower density environments, compared to a control sample drawn from the general ETG population. Around 80 per cent of D-ETGs inhabit the field (compared to 60 per cent of the control ETGs) and less than 2 per cent inhabit clusters (compared to 10 per cent of the control ETGs). Compared to their control-sample counterparts, D-ETGs exhibit bluer ultraviolet-optical colours (indicating enhanced levels of star formation) and an active galactic nucleus fraction that is more than an order of magnitude greater (indicating a strikingly higher incidence of nuclear activity). The mass of clumpy dust residing in large-scale dust features is estimated, using the SDSS $r$-band images, to be in the range $10^{4.5}$--$10^{6.5}$M$_{\odot}$. A comparison to the total (clumpy + diffuse) dust masses -- calculated using the far-infrared fluxes of 15 per cent of the D-ETGs that are detected by the \emph{Infrared Astronomical Satellite (IRAS)} -- indicates that only 20 per cent of the dust is typically contained in these large-scale dust features. The dust masses are several times larger than the maximum value expected from stellar mass loss, ruling out an internal origin. The dust content shows no correlation with the blue luminosity, indicating that it is not related to a galactic scale cooling flow. Furthermore, no correlation is found with the age of the recent starburst, suggesting that the dust is accreted directly in the merger rather than being produced in situ by the triggered star formation. Using molecular gas-to-dust ratios of ETGs in the literature, we estimate
that the median current molecular gas fraction in the \emph{IRAS}-detected ETGs is $\sim$1.3 per cent. Adopting reasonable values for gas depletion time-scales and starburst ages, the median \emph{initial} gas fraction in these D-ETGs is $\sim$4 per cent. Recent work has suggested that the merger activity in nearby ETGs largely involves minor mergers (dry ETG + gas-rich dwarf), with mass ratios between 1:10 and 1:4. If the \emph{IRAS}-detected D-ETGs have formed via this channel, then the original gas fractions of the accreted satellites are between 20 and 44 per cent.
\end{abstract}


\begin{keywords}
galaxies: elliptical and lenticular, cD -- galaxies: evolution -- galaxies: formation -- galaxies: interactions -- galaxies: ISM -- galaxies: peculiar. \vspace{-0.5in}
\end{keywords}


\section{Introduction}
As endpoints of the hierarchical build-up of mass, early-type galaxies (ETGs) possess a detailed fossil record of their progenitors. The stellar populations of ETGs encode the mass assembly history of galaxies over the lifetime of the Universe, and it is important that we gain a comprehensive understanding of their formation and evolution. While their red optical colours \citep*[e.g.][]{bow92,ber03,bel04,fab07}, high alpha-element enhancements \citep*[e.g.][]{tho99,tra00a,tra00b} and obedience of a tight Fundamental Plane \citep*[e.g.][]{jor96,van96,sag97,for98} indicate that the bulk of their stellar populations ($>80$ per cent) are old, the large scatter in the star-formation-sensitive ultraviolet (UV) colours of ETGs since $z\sim 1$ is interpreted as evidence for continuous low-level star formation at late epochs \citep*[][see also \citealt{fuk04,yi05,jeo07,sal10,cro11}]{kav07,kav08,mar09}. A coincidence between blue UV colours and morphological disturbances indicates that this star formation is merger driven \citep{sch90,sch92,kav10,kav11}. Furthermore, the paucity of major mergers at late epochs \citep[see e.g.][]{ste08,jog99,lop10} strongly suggests that the star formation is driven by \emph{minor} mergers \citep{kav09,kav11}. Our current understanding of ETG evolution therefore indicates that their underlying stellar populations are old, having been rapidly built at high redshift ($z>1$), while the evolution at late epochs ($z<1$) is dominated by repeated minor-merger events that contribute $<20$ per cent of their stellar mass at the present day.

Notwithstanding the classical notion of ETGs being dry, passively evolving systems, an extensive literature has developed over the past few decades on the interstellar medium (ISM) in these systems. The dust and gas contents of \emph{very} nearby ETGs have been studied by several authors \citep*[e.g.][]{tub80,haw81,sad85,ber87,ebn88,kna89,gou95,van95,kna96,fab97,tom00,tra01,com07,cal08,you11}. While the galaxy samples and methodologies in these studies are varied, it is clear that the majority of ETGs in the very nearby Universe show evidence of dust \citep[e.g.][]{kna89,van95}, the dust masses being generally inconsistent with scenarios in which the dust is supplied purely by stellar mass loss \citep[e.g.][]{mer98}. Coupled with the frequently observed morphological disturbances and kinematical misalignments between gas and stars in these systems \citep*[e.g.][]{zei90,sag93,ann10,dav11}, it seems likely that a significant fraction of the interstellar matter has an external origin.

The accumulating evidence for widespread minor-merger-driven star formation \citep[e.g.][]{kav09,kav11} makes the study of interstellar matter in ETGs all the more compelling. While the properties of the star formation have been quantified with a reasonable degree of precision, less is known about the fuel that drives this star formation. A study of the ISM of ETGs in modern observational surveys is therefore very desirable. In this paper we present a study of dusty ETGs (D-ETGs), drawn from Data Release 7 (DR7) of the Sloan Digital Sky Survey \citep[SDSS;][]{aba09}, that exhibit prominent, extended dust features in their optical images. In addition to exploring the properties of the dust, the homogeneous nature of the spectrophotometric data from the SDSS allows us to systematically study the star formation, active galactic nucleus (AGN) activity and local environment of the D-ETGs, compare these systems to a statistically meaningful `control' sample drawn from the general ETG population and put some of the results in the literature on a firmer statistical footing. We are also able to explore the properties of the gas in the minor mergers that are responsible for star formation in nearby ETGs, and provide a more complete picture of the star formation activity in massive ETGs at late epochs.

The plan for this paper is as follows. Section~\ref{section:data} describes the selection of D-ETGs from the SDSS and the construction of a control sample to which their properties are compared. In Section~\ref{section:morphologies} we discuss the high incidence of disturbed morphologies observed in the D-ETG sample. In Section~\ref{section:environment} we explore the local environment of these galaxies, and in Section~\ref{section:sf_agn} we employ UV-optical spectrophotometry to study their star formation and AGN activity. We explore the properties and origin of the dust in Section~\ref{section:dust} and conclude by summarising our findings in Section~\ref{section:summary}.


\section{Data}\label{section:data}
\subsection{Sample selection and basic properties}
Galaxy Zoo \citep[GZ;][]{lin08,lin10} is a citizen science project that has used more than 250\,000 members of the general public to morphologically classify $\sim$900\,000 galaxies in the SDSS spectroscopic galaxy sample. The first incarnation of this project [Galaxy Zoo 1 (GZ1)] classified objects in the SDSS galaxy population into broad morphological classes: early types, spirals and mergers. Galaxy Zoo 2 (GZ2), an extension to GZ1 on which this study is based, has explored finer details of galaxy morphologies, such as morphological peculiarities (dust lanes, tidal disturbances, etc.). Note that the GZ objects are taken from the SDSS galaxy catalogue, which does not contain any quasars or type I AGN. Such broad-line objects are identified by a pipeline algorithm which compares the observed SDSS spectrum of individual galaxies to the composite spectrum of \citet{van01}\footnote{http://www.sdss.org/dr7/algorithms/redshift\_type.html} and do not appear in the galaxy catalogue.

\begin{figure*}
\begin{minipage}{172mm}
$\begin{array}{ccc}
\includegraphics[width=2.1in]{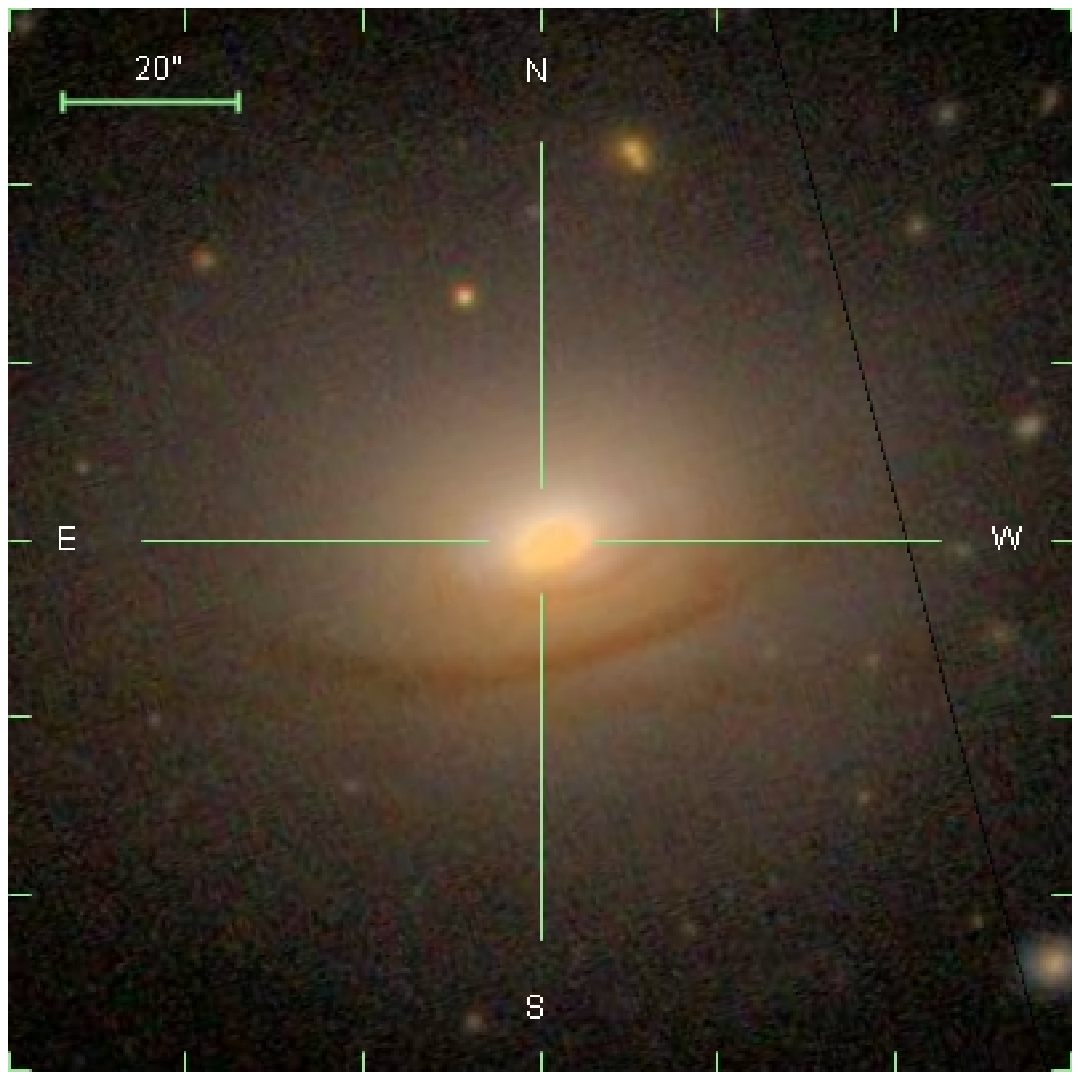} &
\includegraphics[width=2.1in]{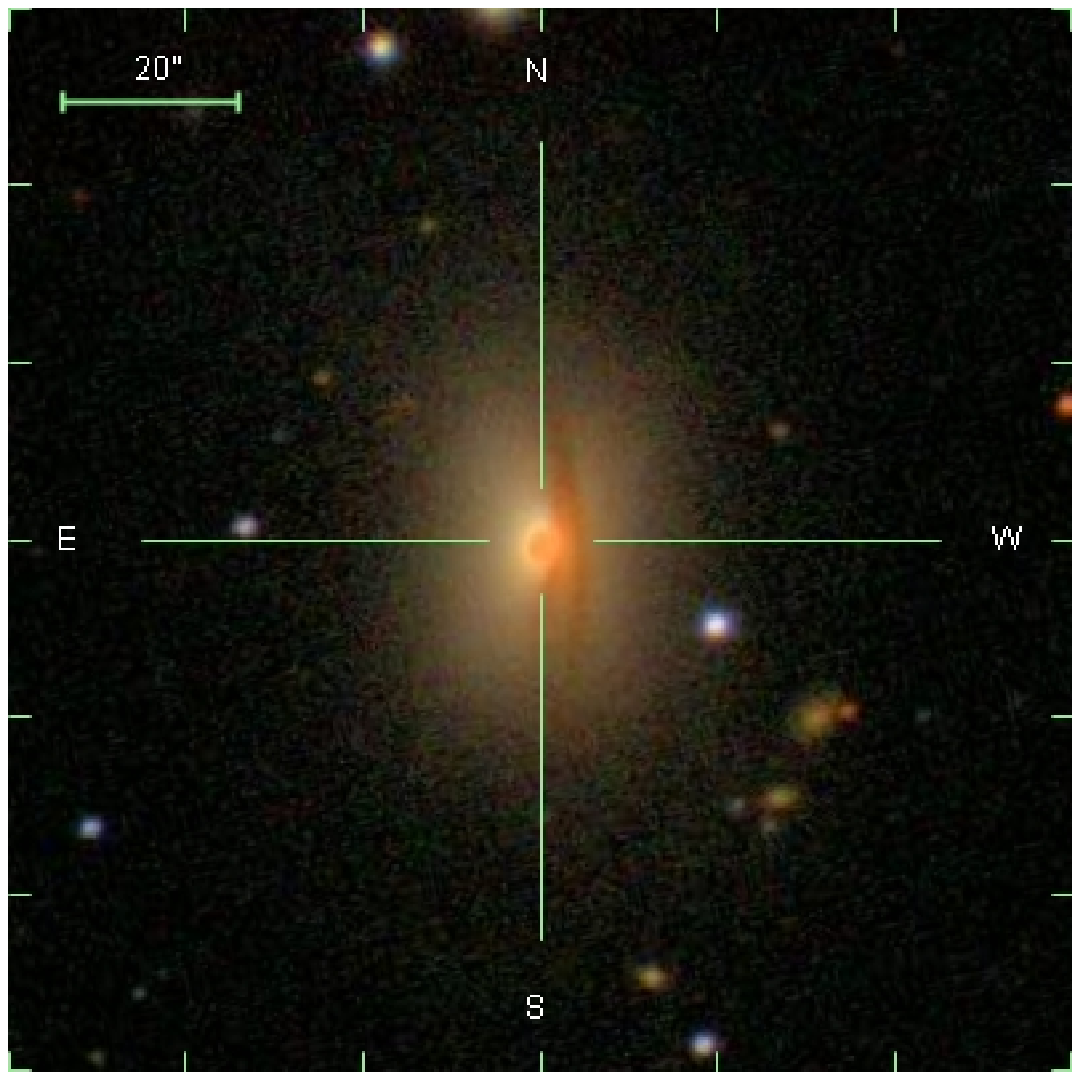} &
\includegraphics[width=2.1in]{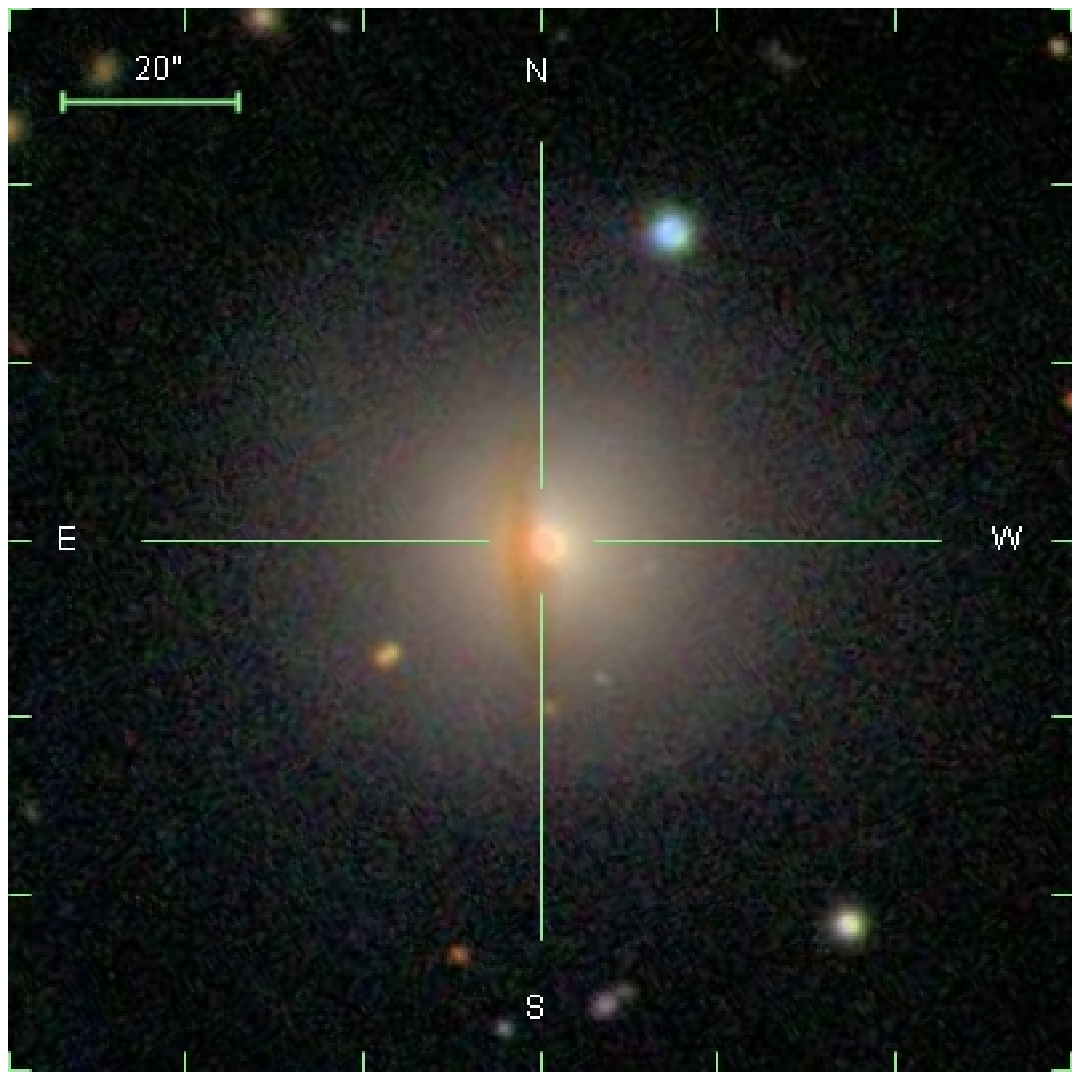}\\
\includegraphics[width=2.1in]{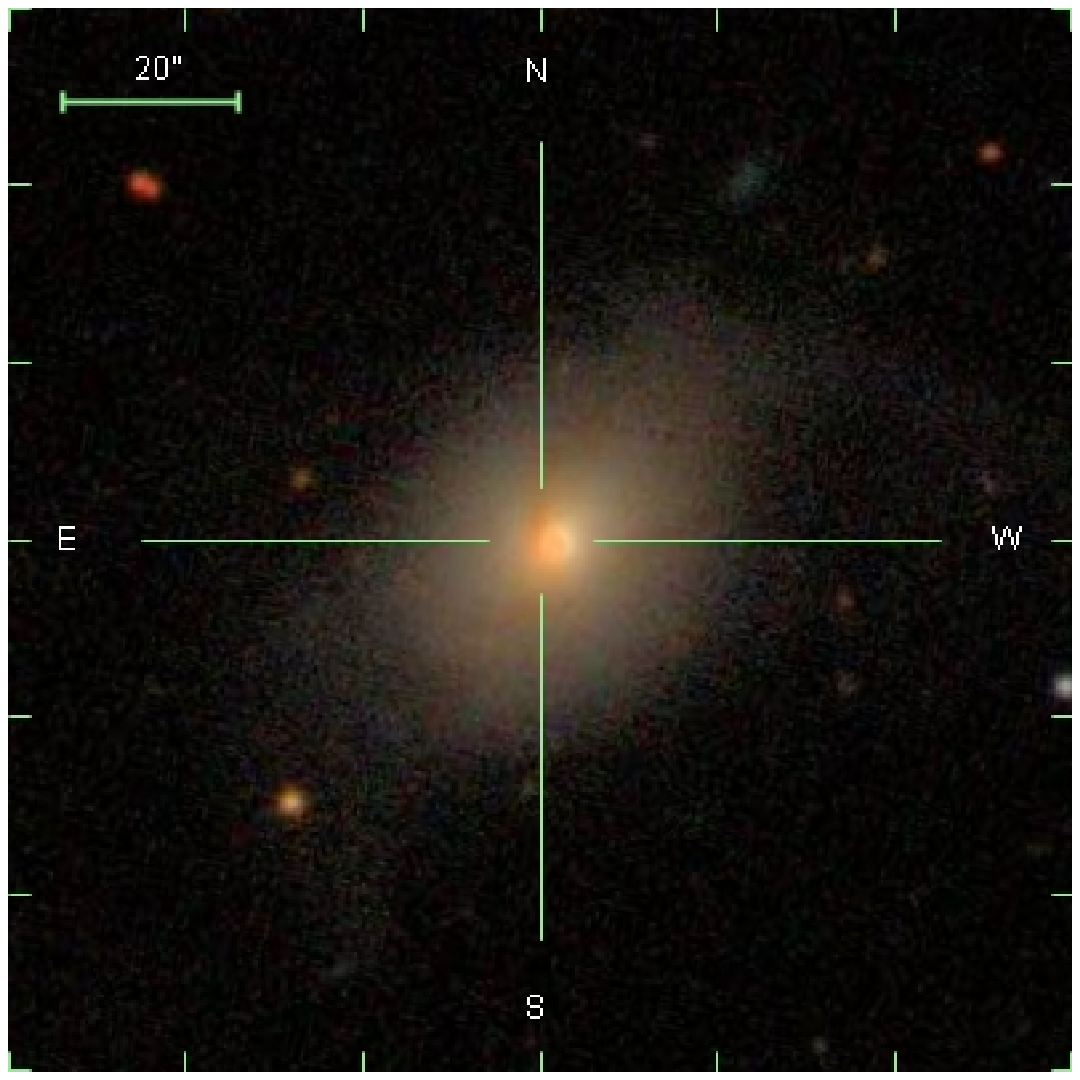} &
\includegraphics[width=2.1in]{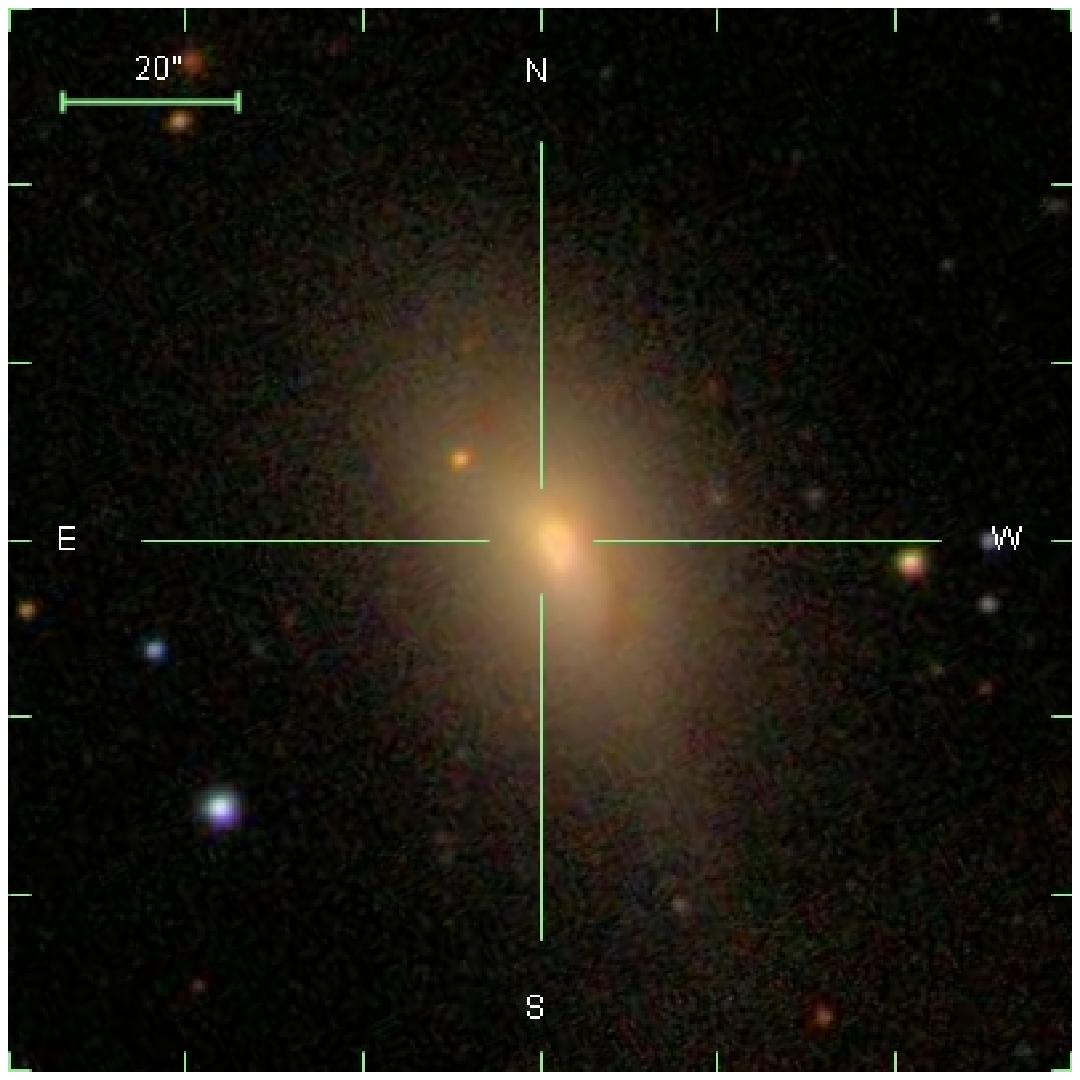} &
\includegraphics[width=2.1in]{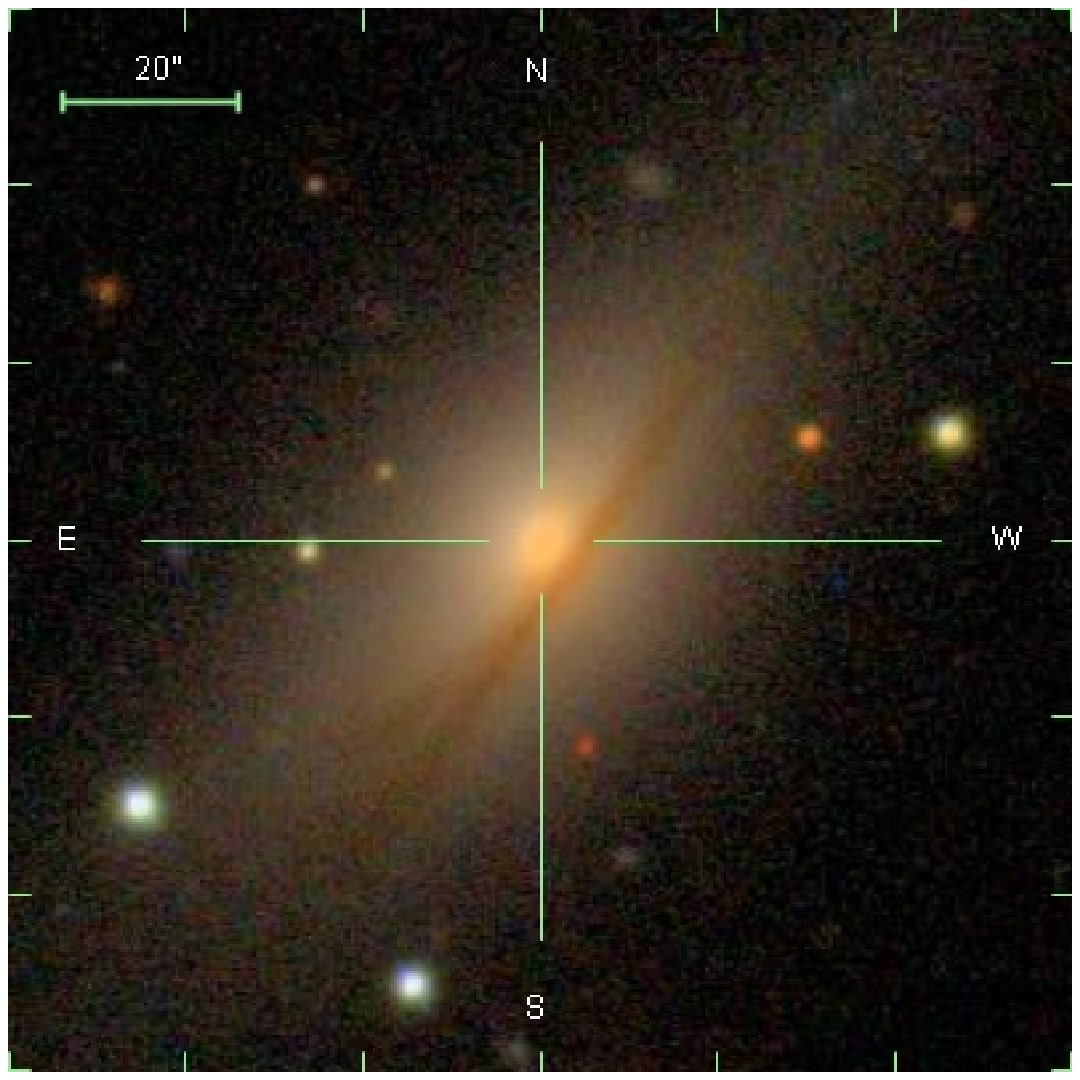}
\end{array}$
\caption{Examples of ETGs with prominent dust features from the sample studied in this paper. Widespread morphological disturbances (shells and tidal tails) are visible in the galaxies. A colour version of this figure is available in the online version of the journal. Note that the faint morphological disturbances may be more readily visible on screen than in printed images.} \label{fig:detg_examples}
\end{minipage}
\end{figure*}

\begin{figure}
$\begin{array}{c}
\includegraphics[width=3.3in]{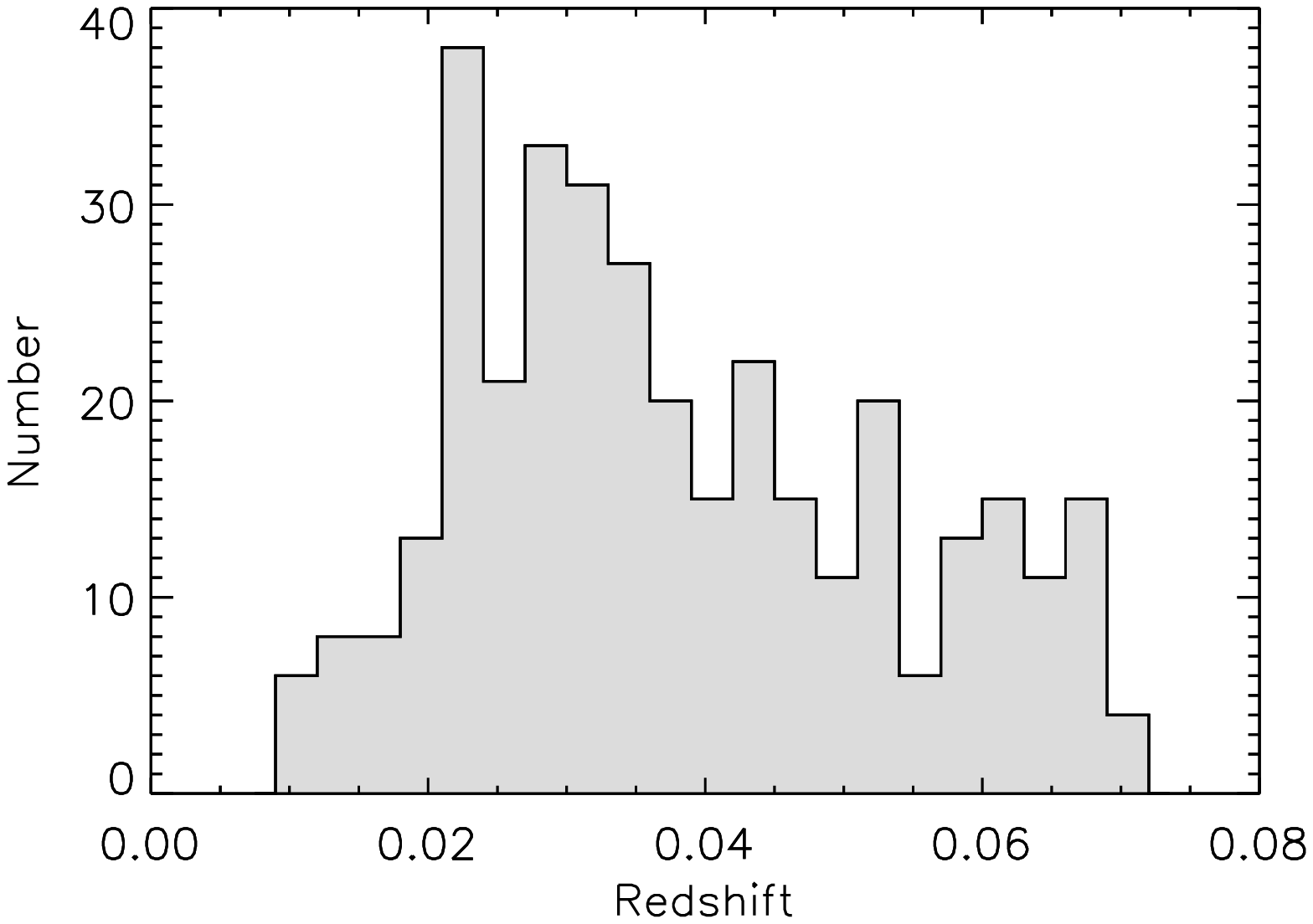}\\
\includegraphics[width=3.3in]{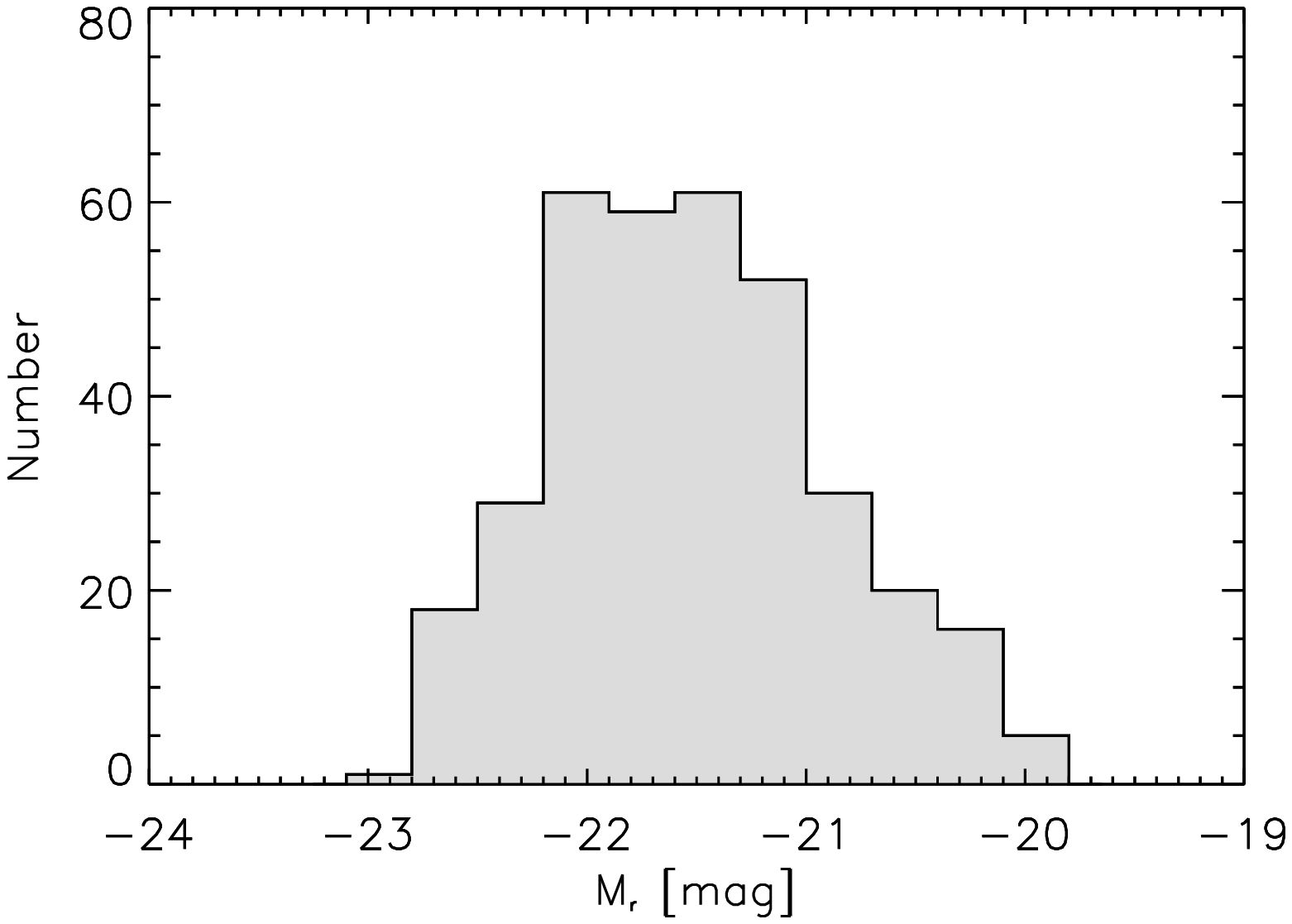}\\
\includegraphics[width=3.3in]{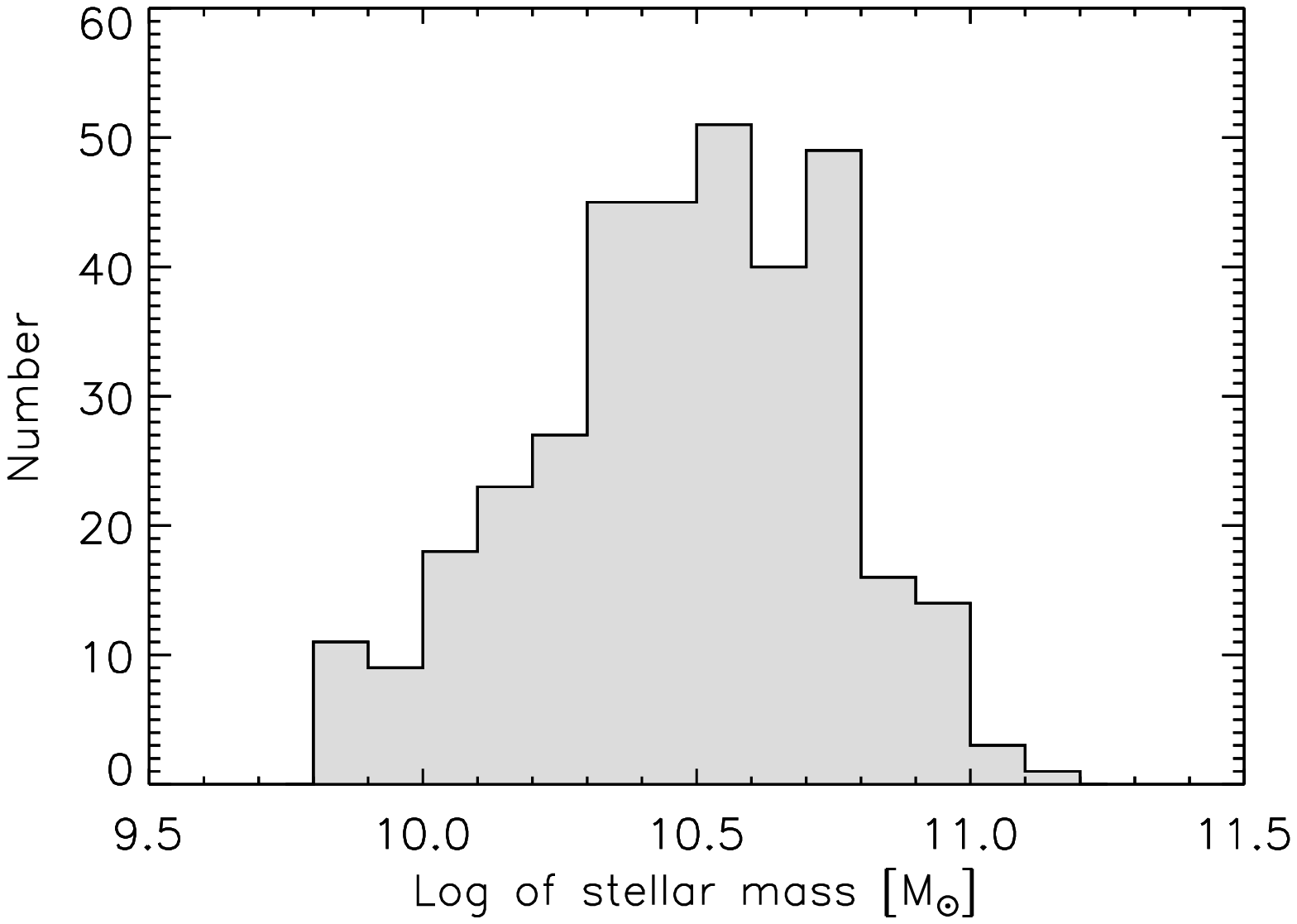}
\end{array}$
\caption{Distributions of redshift (top), $r$-band absolute magnitude (middle) and stellar mass (bottom) of the D-ETG sample in this study. As described in Section~\ref{subsection:control}, the control samples to which these galaxies are compared are matched to the shape of the distributions presented here.}
\label{fig:basic_props}
\end{figure}

\begin{figure}
$\begin{array}{c}
\includegraphics[width=3.5in]{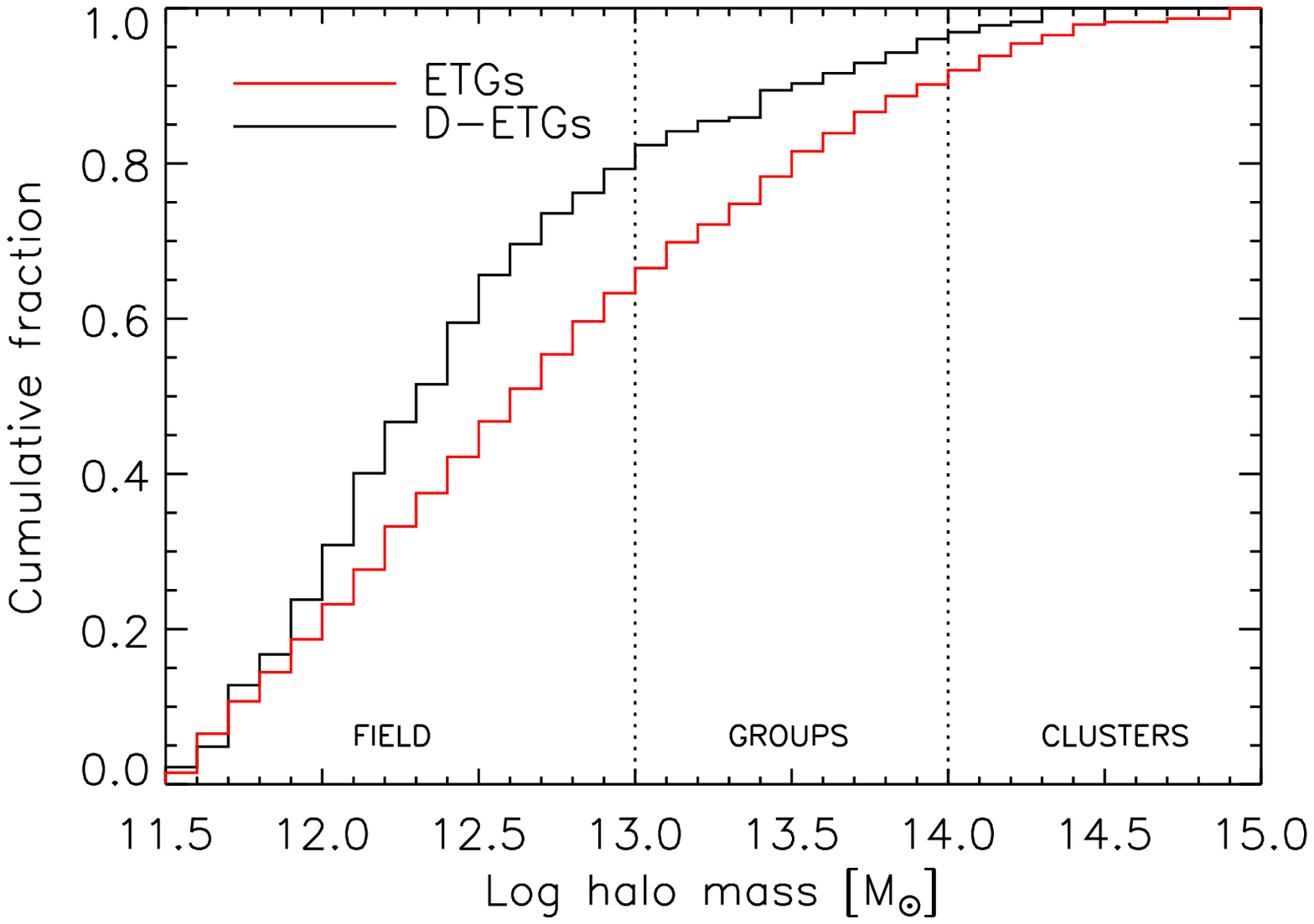}
\end{array}$
\caption{Cumulative distribution of host DM halo masses for D-ETGs and the ETG control sample. Compared to the control sample, D-ETGs preferentially inhabit lower density environments. Almost 80 per cent of the D-ETGs reside in the field (compared to 60 per cent of the control ETG population) and less than 2 per cent of the D-ETGs reside in clusters (compared to 10 per cent of their control-sample counterparts). A colour version of this figure is available in the online version of the journal.}
\label{fig:environments}
\end{figure}

Around 19\,000 GZ2 galaxies in the redshift range $0.01<z<0.1$ are flagged as containing a dust feature by \emph{at least} one GZ2 user. Each galaxy in this sample was visually re-inspected by S.~Kaviraj and Y.~S.~Ting to determine (1) whether the galaxy really has a dust feature, (2) the morphology of the parent galaxy (ETG or non-ETG) and (3) whether the galaxy is morphologically disturbed. After this second layer of visual inspection we decided to restrict this study to the redshift range $z<0.07$, because the unambiguous detection of a dust feature appears to become difficult beyond this redshift. Note that the identification of D-ETGs is not expected to be dependent on galaxy luminosity. As long as the dust is backlit by the galaxy, it is readily visible. This process yields a final sample of reliably classified 352 D-ETGs, which are the subject of this paper. They represent 4 per cent of the ETG population (restricted to the same redshift and magnitude ranges as the D-ETG sample).

Fig.~\ref{fig:detg_examples} shows several typical examples of our D-ETGs. Fig.~\ref{fig:basic_props} shows the basic properties of the D-ETG sample: the redshift, absolute $r$-band magnitude and stellar mass distributions. Stellar masses are calculated using the calibrations of \cite{bel03}. All magnitudes shown in this paper have been $K$-corrected, using the latest version of the public {\texttt{\small KCORRECT}} code \citep{bla03,bla07}.


\subsection{Control sample}
\label{subsection:control}
An important objective of this paper is a systematic comparison between the properties of D-ETGs and those of a control sample of galaxies from the general ETG population. We create a control sample of $\sim$4000 ETGs, using galaxies from GZ2 which have an early-type `vote fraction' greater than 0.7, i.e. more than 70 per cent of the users that classified each of these objects flagged it as an ETG. The control sample is selected to be `dustless', i.e. no GZ user flagged any of these galaxies as having a dust lane. The control sample is then visually inspected to remove late-type contaminants (the contamination rate is very low, around 3 per cent) and flag control ETGs that have morphological disturbances. The control sample is further constructed so that it has the same redshift and magnitude distributions as the D-ETG population shown in Fig.~\ref{fig:basic_props}.


\section{Disturbed morphologies: a merger origin for dusty ETGs}
\label{section:morphologies}
The visual inspection described above indicates that $\sim$65 per cent of D-ETGs show clear morphological disturbances, at the depth of the standard SDSS images. In comparison, the control sample (described in Section~\ref{subsection:control}) shows a disturbed fraction of only around 6 per cent. The fraction of disturbed objects in the D-ETG sample is therefore strikingly high, indicating that D-ETGs are the products of recent interaction events and, in agreement with the literature \citep[e.g.][]{gou94}, that the formation of the dust is likely to be the result of these interactions. Indeed the D-ETGs in this sample are likely to represent objects like Centaurus A \emph{in the making} and are thus ideal test beds for studying the merger process in the low-redshift Universe.


\section{Local environment}
\label{section:environment}
We begin by probing the local environment of our D-ETGs, by cross-matching this sample with the SDSS environment catalogue constructed by \citet{yan07} and \citet*{yan08}, who use a halo-based group finder to separate the SDSS into over 300\,000 structures with a broad dynamic range, from rich clusters to isolated galaxies. This catalogue provides estimates of the masses of the host dark matter (DM) haloes of individual SDSS galaxies, which are related to the traditional classifications of environment (`field', `group' and `cluster'). Cluster-sized haloes typically have masses greater than $10^{14}$M$_{\odot}$, while group-sized haloes have masses between $10^{13}$ and $10^{14}$M$_{\odot}$. Smaller DM haloes constitute what is commonly termed the field.

In Fig.~\ref{fig:environments} we present the cumulative distribution of host DM halo mass for both the D-ETGs and the ETG control sample. We find that, compared to the control ETG population, D-ETGs preferentially inhabit lower density environments. Almost 80 per cent of D-ETGs reside in the field (compared to 60 per cent of the control ETGs) and less than 2 per cent reside in clusters (compared to 10 per cent of their control-sample counterparts). This is consistent with the idea that the D-ETGs are merger remnants, since the high relative velocities in cluster environments make mergers unlikely \citep{bin87}.

\begin{figure}
$\begin{array}{c}
\includegraphics[width=3.5in]{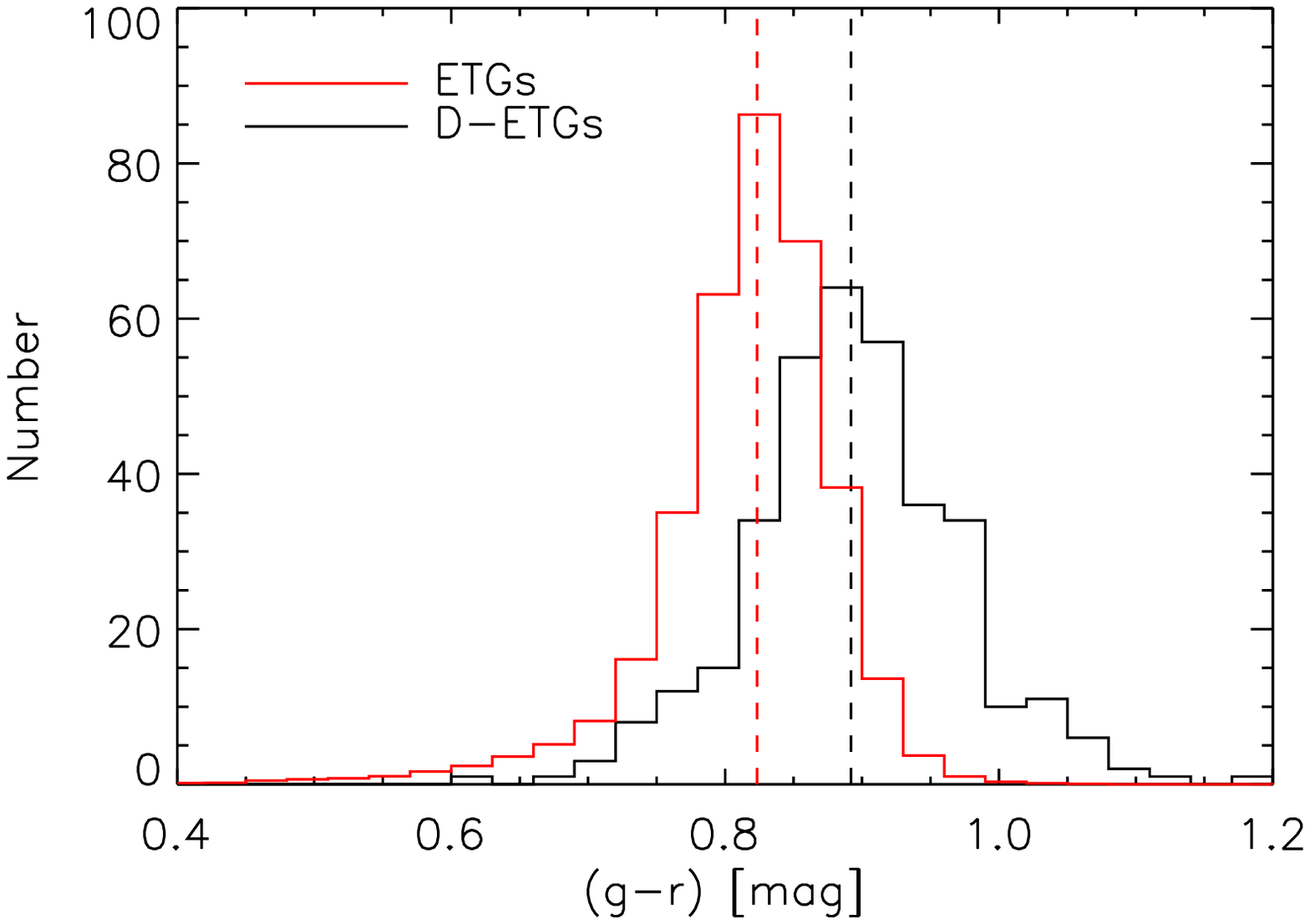}\\
\includegraphics[width=3.5in]{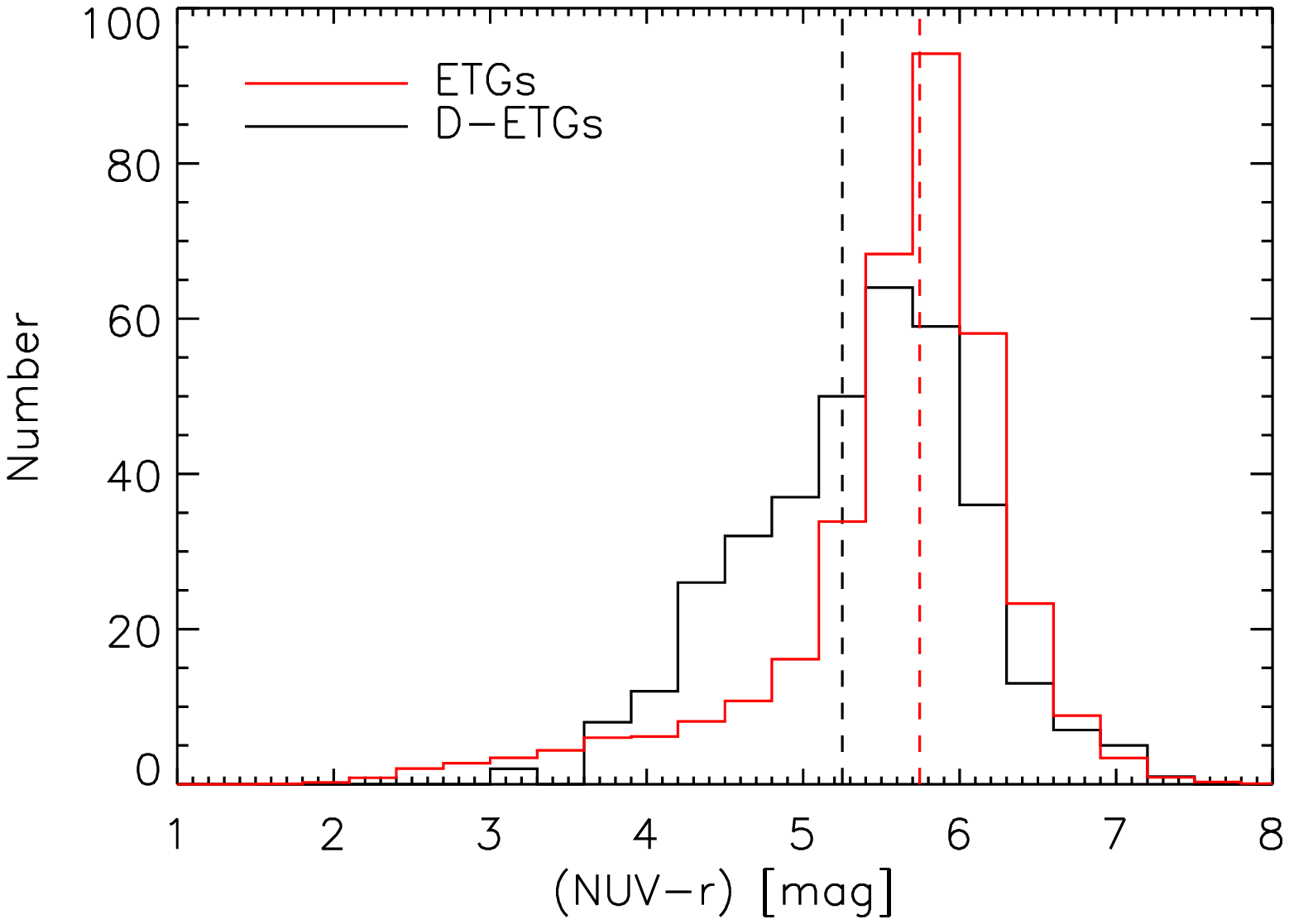}
\end{array}$
\caption{Top: optical $(g-r)$ colour distributions of the D-ETGs and the control ETGs. Unsurprisingly, the D-ETGs are found to be redder than their control counterparts, due to the presence of visibly large amounts of dust. Bottom: $($NUV$-r)$ colour distributions of the D-ETGs and the control ETGs. The D-ETGs are bluer in the UV-optical colour, presumably due to recent star formation (see text for details). Median values are shown using the dashed vertical lines. A colour version of this figure is available in the online version of the journal.}
\label{fig:colours}
\end{figure}


\section{Star formation and AGN activity}
\label{section:sf_agn}
Fig.~\ref{fig:colours} shows the $(g-r)$ and $($NUV$-r)$ colour distributions of the D-ETGs compared to those of their control-sample counterparts. The optical $g$ and $r$ magnitudes are taken from the SDSS \citet{yor00}. The NUV magnitudes are from the \emph{Galaxy Evolution Explorer (GALEX)} space telescope \citep{mar05}. The \emph{GALEX} NUV filter is centred at $\sim$2300 \AA \citep{mor07}. We use the recommended total magnitudes in the AB systems for both the SDSS and \emph{GALEX} photometry \citep[][]{sto02,mor07}.

We find that the D-ETGs are redder than their control counterparts in the optical $(g-r)$ colour, which is consistent with the visibly high levels of dust present in these systems. Nevertheless, the D-ETG population is bluer than the control sample in the $($NUV$-r)$ colour, with a median offset of $\sim$0.5 mag. Given that the UV wavelengths are several times more sensitive to young stars than the optical wavelengths \citep[see e.g. fig. 1 in][]{kav09}, the bluer UV-optical colours are likely to be driven by recent or ongoing star formation in the D-ETG population. It is important to note that the large-scale structure of the dust in the D-ETGs is clumpy and does not entirely obscure the galaxy core (see Fig.~\ref{fig:detg_examples}), where the young stars (which dominate the UV flux) are typically concentrated \citep[e.g.][]{fer09,jeo09,pei10,suh10}.

We proceed by probing the dominant gas ionization mechanism in the D-ETG sample. The presence of dust implies the availability of gas which, in turn, makes the presence of star formation and nuclear activity in these galaxies likely. We use the standard optical emission-line ratio analysis of \citet[][see also \citealt*{bal81,vei87,kau03}]{kew06} which exploits the [N\,{\sc ii}]/H$\alpha$ and [O\,{\sc iii}]/H$\beta$ ratios to probe the presence of type II AGN in our D-ETG sample. The optical emission-line fluxes are calculated from the SDSS spectrum of each object using the publicly available code {\texttt{\small GANDALF}} \citep{sar06}. Note that the SDSS spectra are measured in 3-arcsec fibres ($\sim$ 4 kpc at $z=0.035$, the median redshift of our D-ETG sample) centred on the galaxy. Galaxies in which all four emission lines are detected with a signal-to-noise ratio greater than 3 are classified as either `star forming', `composite' (i.e. hosting both star formation and AGN activity), `Seyfert' or `LINER'. Galaxies without a detection in all four lines are classified as `quiescent'. The fraction of galaxies in each emission-line class is listed in Table~\ref{table:agn}, for both the D-ETG and control samples.

We find that the fractions of D-ETGs classified as star forming and composite are, respectively, factors of 3 and 8 larger than that for the control ETGs. This suggests enhanced levels of star formation in the D-ETG population compared to its control counterpart, consistent with the UV-optical colour analysis above. The fraction of LINERs in the D-ETGs is also greater by a factor of 3, plausibly due to LINER-like emission from merger-driven shocks or evolved stellar populations (such as post-asymptotic giant branch stars) that are associated with the star formation \citep[e.g.][]{sar10}. Table~\ref{table:agn} indicates that, in addition to enhanced star formation and LINER emission, the D-ETGs exhibit a strikingly higher Seyfert fraction, more than an order of magnitude larger than in their control-sample counterparts. The fraction of quiescent galaxies is correspondingly an order of magnitude smaller. Therefore, the presence of prominent dust features in an ETG significantly increases its chances of hosting both more star formation and more nuclear activity, with nuclear activity being the dominant gas ionization mechanism in these systems.


\section{Properties of the dust and associated gas}
\label{section:dust}

\subsection{Clumpy dust}
\label{subsection:clumpy_dust}

\begin{table}
\begin{center}
\caption{Comparison of star formation and AGN activity, derived using optical emission-line ratio diagnostics, in the D-ETGs and their control-sample counterparts.}
\begin{tabular}{lrrr}\hline

                  & D-ETGs     & Control ETGs  & $f$(D-ETG)/$f$(Ctrl)\\
                  & (per cent) & (per cent)    &                     \\\hline
Star forming      & 7.2        & 2.1           & 3.4                 \\
Composite         & 22.8       & 2.8           & 8.1                 \\
Seyfert           & 19.9       & 1.7           & 11.7                \\
LINER             & 37.5       & 10.4          & 3.6                 \\
Quiescent         & 12.8       & 83.0          & 0.2                 \\\hline

\end{tabular}
\label{table:agn}
\end{center}
\end{table}

The dust in D-ETGs is comprised of both clumpy structures, which are easily visible in the optical images, and diffuse dust, which is intermixed with the stars and is essentially invisible in the optical imaging. We begin by calculating the dust contained within the clumpy features ($M^{\textrm{CL}}_{\textrm{d}}$), following a method similar to that employed by several previous studies \citep[e.g.][]{sad85,van95,tra01}. The clumpy dust mass is given by
\begin{eqnarray}
M^{\textrm{CL}}_{\textrm{d}} = \langle A_V \rangle \Sigma
\Gamma_V^{-1},
\label{equation:extinction}
\end{eqnarray}

\noindent where $\langle A_V \rangle$ is the mean $V$-band extinction in magnitudes, $\Sigma$ is the total area of the dust features and $\Gamma_V$ is the mass absorption coefficient.\footnote{The right-hand side of equation~(\ref{equation:extinction}) is derived as follows. If $\delta$ is the area of a single pixel and $A_i$ is the extinction in pixel $i$, then: $M^{\textrm{CL}}_{\textrm{d}}=[\delta A_1 + \delta A_2 + \delta A_3 + \cdots + \delta A_n]\Gamma_V^{-1}$. Since the total area $\Sigma \equiv n \delta$, $M^{\textrm{CL}}_{\textrm{d}}= (\Sigma/n) [A_1 + A_2 + A_3 + \cdots + A_n]\Gamma_V^{-1}=\langle A \rangle \Sigma \Gamma_V^{-1}.$}

\begin{figure}
$\begin{array}{c}
\includegraphics[width=3.5in]{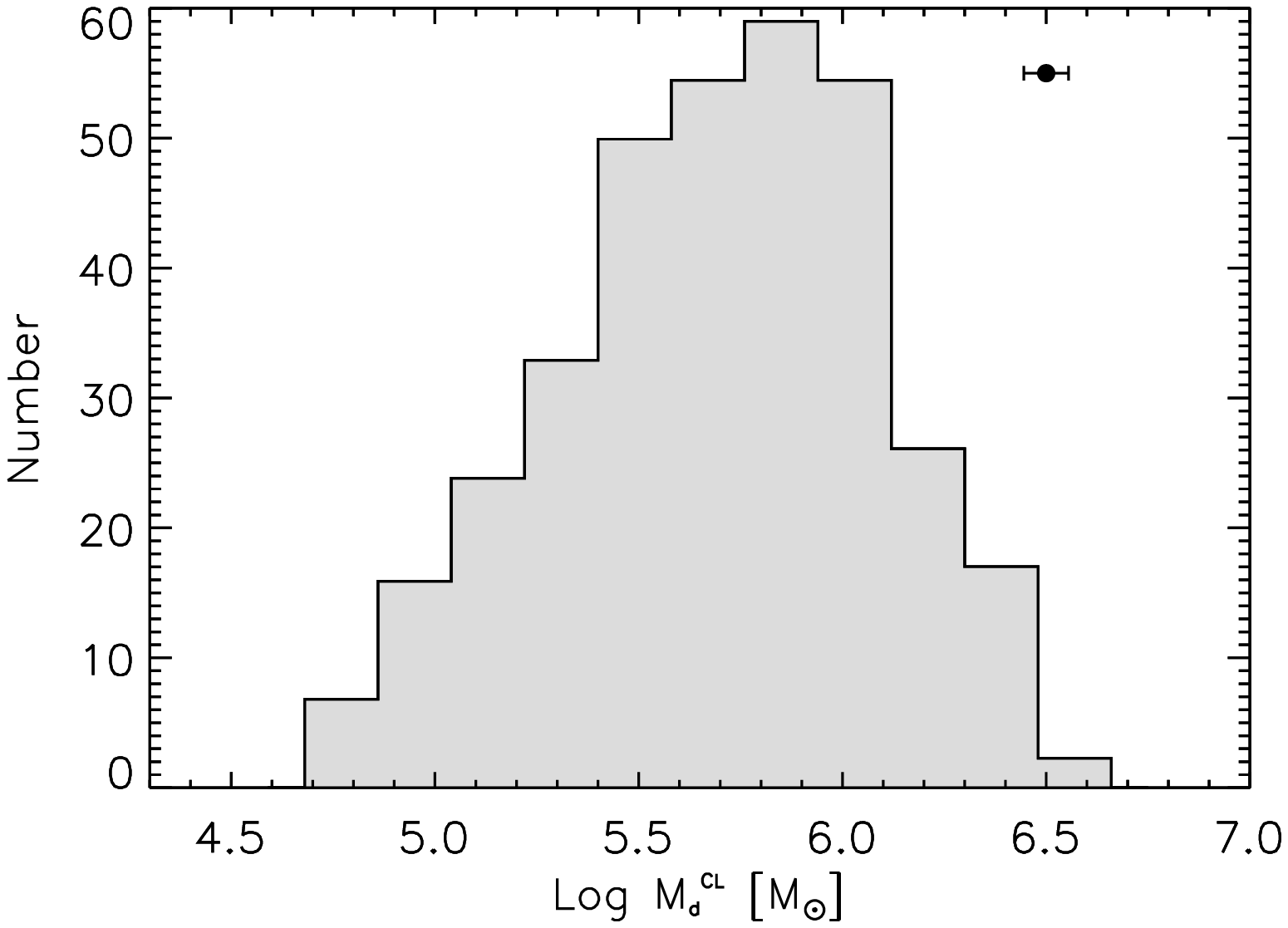}\\
\includegraphics[width=3.5in]{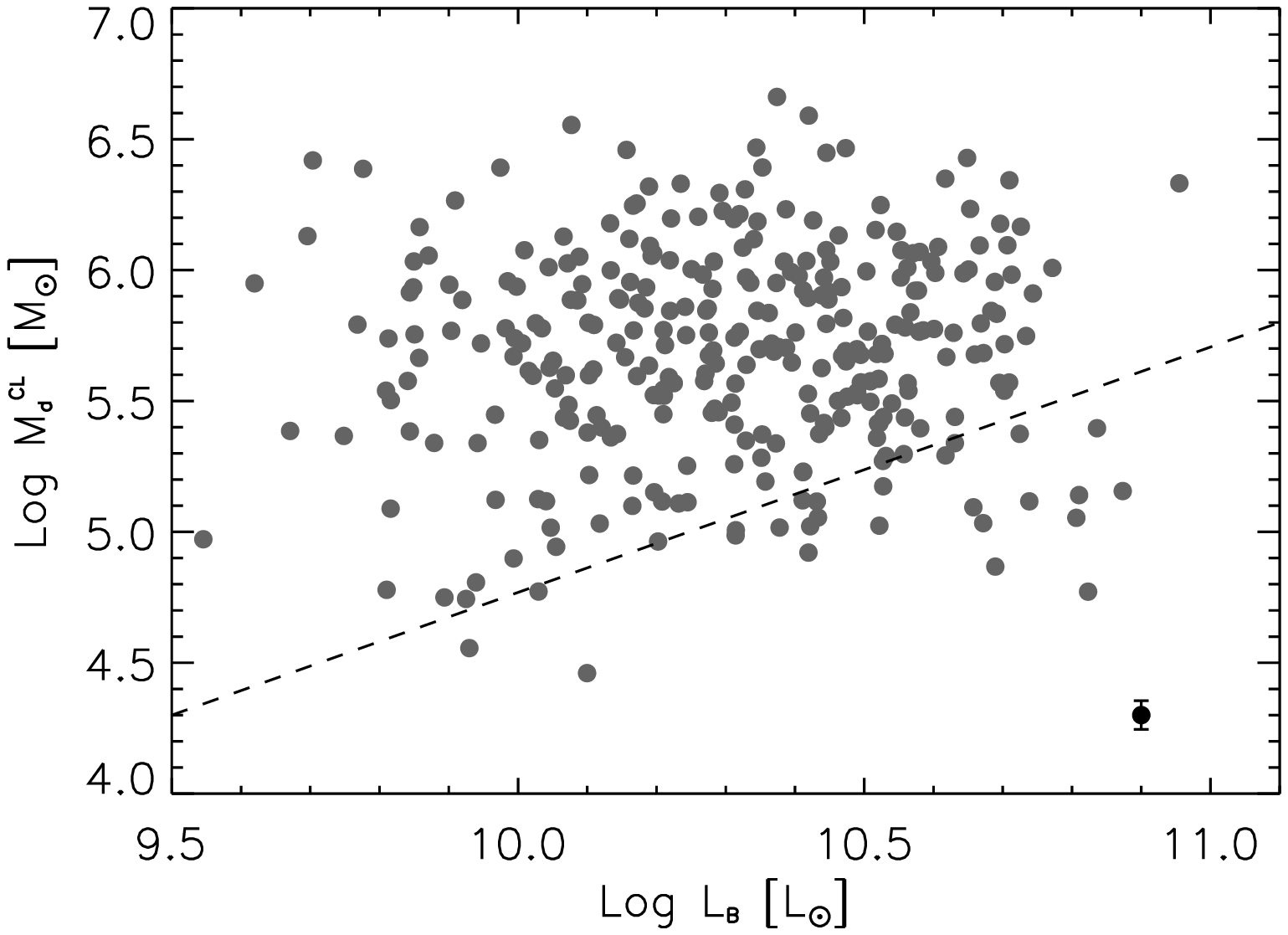}\\
\includegraphics[width=3.5in]{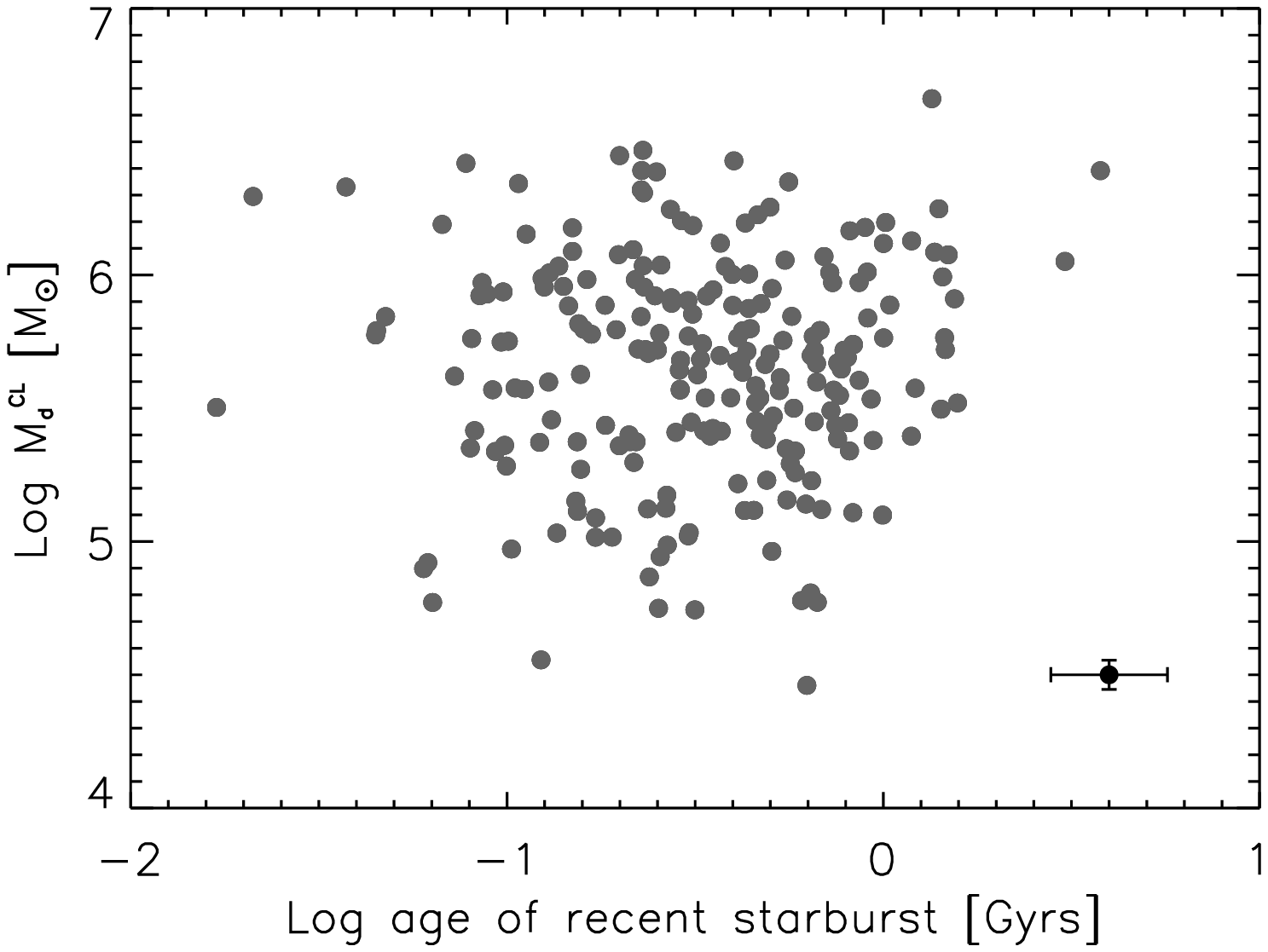}
\end{array}$
\caption{Top: distribution of clumpy dust mass ($M^{\textrm{CL}}_{\textrm{d}}$) in the D-ETGs, derived from their optical $r$-band images (see text for details). Middle: clumpy dust mass versus blue luminosity of the D-ETGs. Bottom: clumpy dust mass versus age of the recent starburst, estimated from UV-optical colours. Note that, while the formal error on $M^{\textrm{CL}}_{\textrm{d}}$ is $\sim 0.055$ dex (driven by the uncertainties in $A_V$), the uncertainties in the assumed dust distribution induces an additional error of up to 0.3 dex (see text for details).}
\label{fig:dust_properties}
\end{figure}

Following these previous studies, $A_V$ is measured as follows. Ellipses are first fitted to the isophotes of the SDSS $r$-band images of each galaxy, using the {\texttt{\small IRAF}} task \texttt{ELLIPSE}, with the dust features masked out. An absorption map is then created, using the ratio of the flux in the observed
image ($F_{\textrm{obs}}$) to the flux in the model fit ($F_{\textrm{mod}}$). The optical depth is $\tau = -\textrm{ln}(F_{\textrm{obs}}/F_{\textrm{mod}})$ and $A_r=1.09\tau$ \citep{tra01}. We convert $A_r$ to $A_V$ using the Galactic dust law: $A_V = 1.33 A_r$ \citep{tra01}. Note that the quality of the \texttt{ELLIPSE} fits are unaffected by the faint tidal features that are found on the outskirts of the D-ETGs.

Several studies have shown that the extinction curves in ETGs with dust lanes are similar to their Galactic counterpart \citep*[see e.g.][]{sad85,gou94,sah98,fin08}, which indicates that the properties of the dust in these systems are similar to that in the Galaxy. Therefore, it appears reasonable to employ the Galactic value for the mass absorption coefficient (i.e. $\Gamma_V = 6 \times 10^{-6}$ mag kpc$^{2}$ M$_{\odot}^{-1}$) although we note that a more accurate calculation requires a measurement of $\Gamma_V$ in our individual D-ETGs. The formal error on $M^{\textrm{CL}}_{\textrm{d}}$ (driven, in this case, only by the error in $\langle A_V \rangle$) is calculated using the standard error propagation formula. We note that the errors in other quantities that appear later in this study (e.g. $M^{\textrm{IRAS}}_\textrm{d}$ and $g_\textrm{i}$) are also calculated using the standard error formula.

While the formal error on $M^{\textrm{CL}}_{\textrm{d}}$ is only $\sim$0.055 dex, this derivation assumes that the dust is placed in a simple foreground screen in front of the light source; however, in actual fact the dust is embedded in the galaxy. The use of this simplifying assumption induces an additional uncertainty of up to $\sim$0.3 dex in the dust mass calculation \citep[e.g.][]{tom00,tra01}. The values derived here (see the top panel of Fig.~\ref{fig:dust_properties}) are in the range $10^{4.5}$--$10^{6.5}$M$_{\odot}$, in good agreement with the range of clumpy dust masses derived using similar techniques in the literature \citep[10$^3$--10$^7$M$_{\odot}$;][]{sad85,gou95,van95,tom00,tra01,whi08}.

In the middle panel of Fig.~\ref{fig:dust_properties}, we plot the clumpy dust mass of the D-ETGs against their total blue luminosity ($L_B$). $L_B$ is calculated from the absolute (total) $B$-band magnitude, after transforming from the absolute SDSS (total) $g$-band magnitude using the relation $m_B =m_g + 0.3$ \citep{rod06}. We use $L_B$ to facilitate direct comparison with previous studies \citep[e.g.][]{mer98}. The dashed line indicates the maximum expected contribution to the dust content from stellar mass loss alone \citep*[e.g.][]{fab76,kna92}, assuming that the dust is depleted by sputtering \citep*[e.g.][]{spi78,dra79a,dra79b,dwe86,dwe90} over a maximal destruction time-scale of $\sim10^{7.5}$ yr \citep{fab76}. In agreement with previous studies \citep[e.g.][]{gou94,mer98}, we find that (1) D-ETGs typically have clumpy dust masses that are well in excess of the maximum equilibrium mass expected from stellar mass loss alone and (2) there is no correlation between the optical dust masses and $L_B$. It is worth noting that $L_B$ shows a correlation with the total X-ray luminosity ($L_X$) of ETGs, which traces their hot gas content \citep*[e.g.][]{can87,mat03,mul10}. The lack of a correlation between the clumpy dust mass and $L_B$ suggests that the clumpy dust is not produced through cooling from the hot gas reservoirs that may be present in these ETGs \citep*[see also][]{for91,bre92}. Recall also that the D-ETGs studied in this paper do not typically inhabit clusters, so that cluster-scale cooling flows are not relevant to these systems.

The dust properties derived above indicate that the clumpy dust content of the D-ETGs is not created by internal processes such as stellar mass loss or gas cooling. Together with the high incidence of morphological disturbances observed in the D-ETG sample (see Section~\ref{section:morphologies}), this indicates that the clumpy dust observed in these galaxies is likely to be associated with the recent merger events that created these objects.

Since star formation is clearly enhanced in D-ETGs, it is worth exploring whether the dust is produced in situ by this star formation or whether it is directly accreted in the mergers. In the bottom panel of Fig.~\ref{fig:dust_properties} we plot the clumpy dust mass of D-ETGs against the age of the recent starburst ($T$), which has been estimated using the UV-optical colour \citep{kav10}:
\begin{eqnarray}
\log T \hspace{0.05in}\textnormal{[Gyrs]} \sim 0.6^{\pm0.03}\times[(\mbox{NUV}-u)-(g-z)-1.73^{\pm0.03}].
\label{equation:age}
\end{eqnarray}

\noindent Briefly, \citep[see sections 4.2.1 and 4.3 in][for details]{kav10}, $T$ is one of several star formation history (SFH) parameters derived for individual ETGs by fitting their \emph{GALEX}/SDSS UV-optical photometry to a library of synthetic photometry, generated using a large collection of model SFHs. The model SFHs assume two instantaneous bursts, the first fixed at $z=3$ (since the bulk of the stars in ETGs are old), while the second (which represents the recent star formation episode) is allowed to vary in age ($T$) and mass fraction (realistic ranges of dust and metallicity are applied to the model SFHs). \citet{kav10} extract marginalized values for the free parameters including $T$ and find a strong correlation between the marginalized values of $T$ and the relatively dust-insensitive double colour $[($NUV$-u)-(g-z)]$. This relation (equation~\ref{equation:age}) allows us to estimate the age of the recent starburst in ETGs from the observed value of $[($NUV$-u)-(g-z)]$.

Fig.~\ref{fig:dust_properties} shows that the clumpy dust mass does not correlate with $T$. The apparent decoupling between the age of the recent (merger-induced) star formation episode and the clumpy dust content in D-ETGs favours a scenario in which \emph{the dust is directly accreted during the merger and not created in situ by the triggered star formation.}


\subsection{Diffuse dust and the total dust content}
Since the diffuse dust mass in the ISM cannot be estimated from the optical images, the clumpy dust masses calculated above are \emph{likely to be lower
limits} to the total dust masses. It is thus desirable to quantify the total (diffuse + clumpy) dust content of these systems, at least to gain an understanding of how much dust is not contained in the clumpy features visible in the optical images. A useful route to estimating the total dust content is to use the far-infrared spectrum. Around 15 per cent of our D-ETGs have infrared data from the \emph{Infrared Astronomical Satellite (IRAS)}, which we use to estimate the `total' dust mass in these galaxies. It should be noted, however, that \emph{IRAS} is insensitive to very cold dust, with a temperature less than $\sim$20 K \citep[e.g.][]{hen97}. Thus, while the \emph{IRAS} fluxes are a much better estimator of the total dust mass than the optical images, more accurate estimates require imaging by new instruments such as \emph{Herschel}, that probe the far-infrared spectrum beyond 100 $\mu$m.

\begin{figure}
$\begin{array}{c}
\includegraphics[width=3.5in]{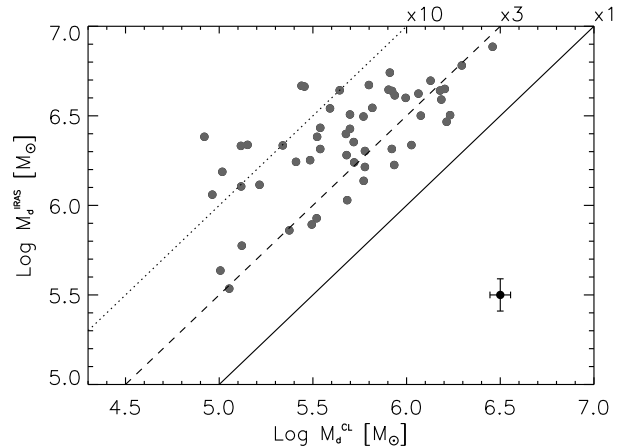}
\end{array}$
\caption{Comparison of the clumpy dust masses ($M^{\textrm{CL}}_{\textrm{d}}$), derived from the optical $r$-band images, with the dust masses derived from \emph{IRAS} fluxes $M^{\textrm{IRAS}}_{\textrm{d}}$. The \emph{IRAS}-derived dust masses are typically a factor of 4.5 larger than their clumpy counterparts. Note that the error bar in $M^{\textrm{IRAS}}_{\textrm{d}}$ is the error from equation (\ref{equation:dust}).}
\label{fig:iras_dust}
\end{figure}

Following past studies \citep[e.g.][]{gou95,tra01}, the \emph{IRAS}-derived dust masses ($M^{\textrm{IRAS}}_\textrm{d}$) are calculated using the formula
\citep[see e.g.][]{sch82,hil83,row86,kwa92}
\begin{eqnarray}
M^{\textrm{IRAS}}_\textbf{d} = 5.1 \times 10^{-11}S_{\nu}D^2\lambda_{\nu}^4 [e^{(1.44\times 10^4/\lambda_{\nu}T_\textrm{d})}-1] \hspace{0.05in} M_{\odot},
\label{equation:dust}
\end{eqnarray}

\noindent where the dust temperature ($T_\textrm{d}$) is estimated using the \emph{IRAS} flux ratio $S_{60}/S_{100}$ and an emissivity law that varies as $\lambda^{-1}$, $S_{\nu}$ is the \emph{IRAS} flux density in mJy at wavelength $\nu$, $D$ is distance to the galaxy in Mpc and $\lambda_{\nu}$ is the wavelength in microns. We use $\nu=100 \hspace{0.03in} \mu$m for our calculations. Fig.~\ref{fig:iras_dust} shows the \emph{IRAS}-derived dust masses plotted against the clumpy dust masses, for the subset of D-ETGs that have \emph{IRAS} detections at 60- and 100-$\mu$m. The \emph{IRAS} fluxes are taken from the Imperial IRAS-FSC Redshift Catalogue \citep{wan09}. The median discrepancy between the clumpy dust mass and the \emph{IRAS} dust mass is a factor of 4.5. In other words, only $\sim$20 per cent of the dust is typically contained in clumpy features. Nevertheless, the \emph{IRAS} dust masses show the same lack of correlation with blue luminosity and starburst age as the clumpy dust masses do, leaving our conclusions above unchanged.

\begin{figure}
$\begin{array}{c}
\includegraphics[width=3.5in]{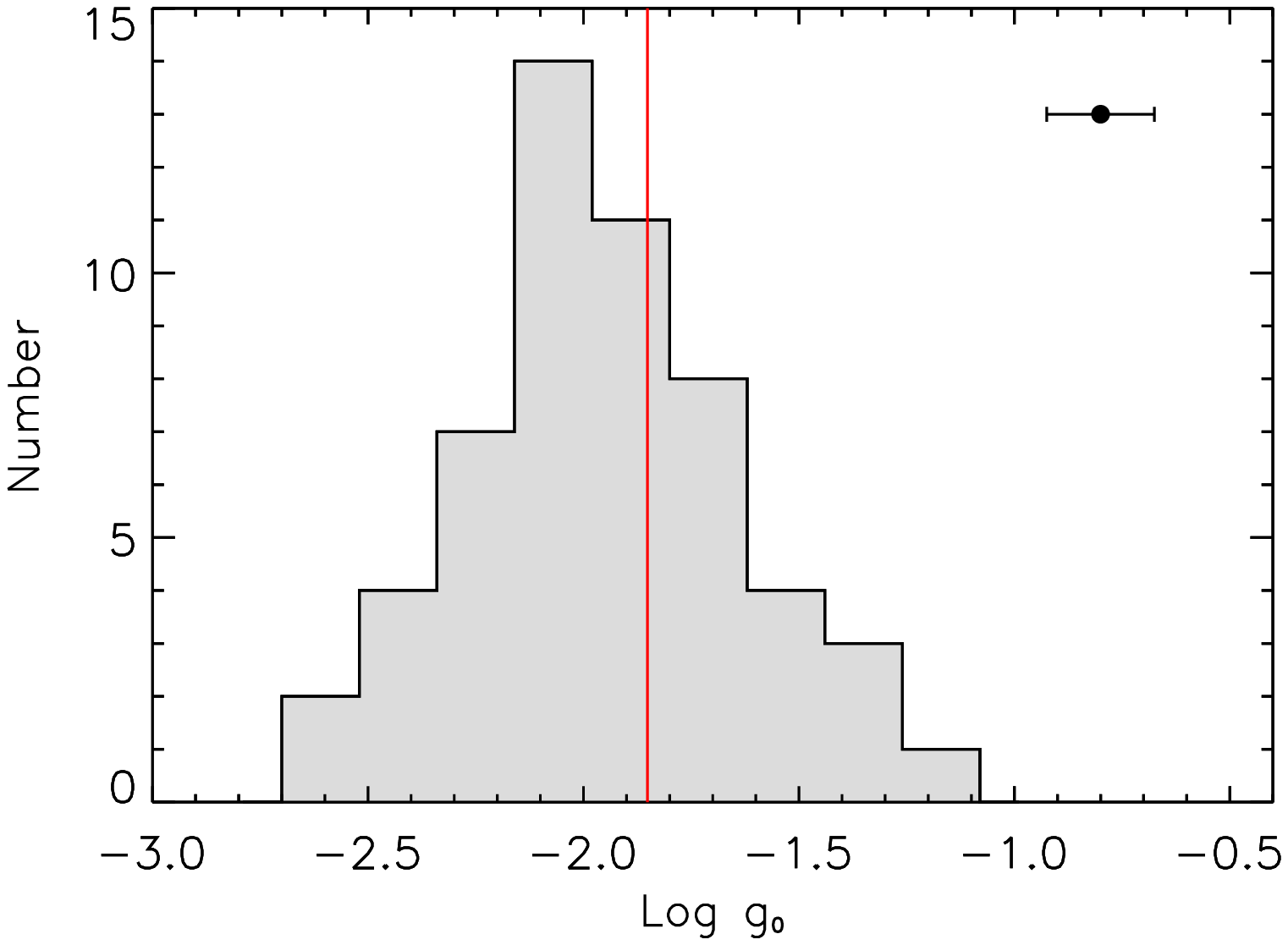}\\
\includegraphics[width=3.5in]{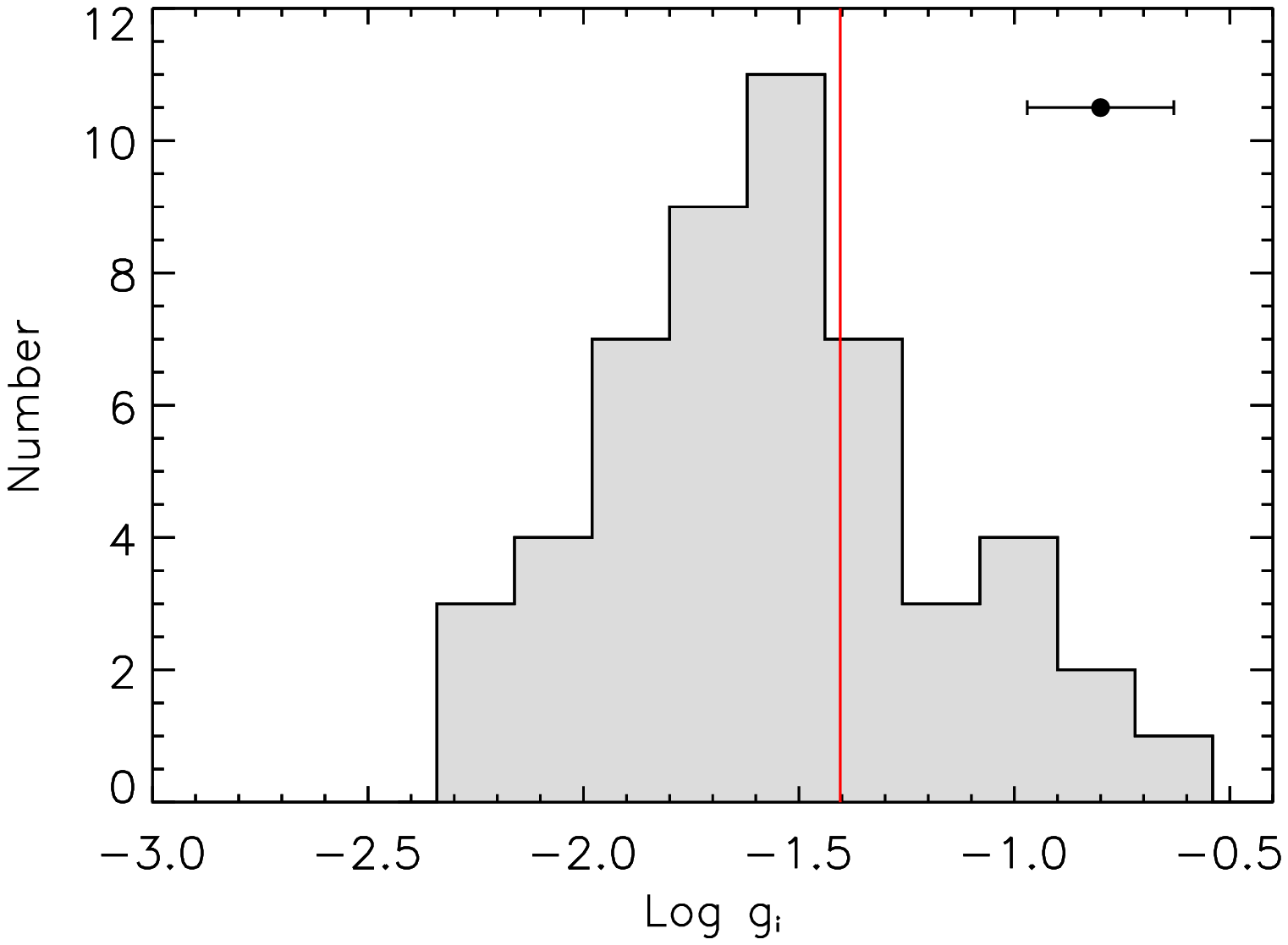}
\end{array}$
\caption{Derived current (top) and initial (bottom) molecular gas fractions (i.e. molecular gas mass divided by the stellar mass) in the subset of D-ETGs that are detected by \emph{IRAS} (see text for details). Median values of the distributions are indicated using solid (red) lines.} \label{fig:gf}
\end{figure}


\subsection{Cold molecular gas}
Given the close correspondence between dust and cold gas \citep[e.g.][]{kna99}, it is instructive to explore the properties of the cold gas reservoirs of our D-ETGs. We choose to focus on \emph{molecular} gas (i.e. $\textnormal{H}_\textrm{2}$, derived from CO observations) because of its relevance to star formation \citep[e.g.][]{won02,sch04,kru05,ler08} and the better availability of molecular gas data on ETGs in the literature compared to its atomic counterpart (H\,{\sc i}). We note first that previous studies have typically assumed that molecular gas-to-dust ratios (G/D values) in ETGs are similar to that in the Galaxy \citep[$\sim$150; see e.g.][]{spi78,hil83,dra84}. However, if the dust is directly accreted during recent mergers (c.f. Section~\ref{subsection:clumpy_dust}), then the Galactic G/D is inappropriate for all ETGs, unless the accreted companions are all Milky Way like. In fact, recent work \citep[e.g.][]{kav09,kav11,cro11} has shown that a large proportion of the merger activity experienced by nearby ETGs involves interactions with small gas-rich companions (rather than Milky Way like systems), implying that the Galactic G/D value may not be appropriate for our D-ETG sample.

To obtain a more realistic conversion between dust and molecular gas for our D-ETGs, we appeal to measurements of molecular G/D values in ETGs that are available in the literature. For 76 ETGs studied by \citet*{wik89,thr89,gor91,lee91,wan92} and \citet{wik95}, the average molecular G/D is $1190 \pm 410$. Thus, the molecular G/D values appropriate for our D-ETGs are likely to be almost an order of magnitude higher than the value in the Galaxy. We note here that the measured G/D values in ETGs are indeed around the values typical of late-type dwarfs \citep[e.g.][]{you89}, further supporting the view that the interactions that produce the D-ETGs probably involve the accretion of gas-rich satellites. In the top panel of Fig.~\ref{fig:gf}, we use the \emph{IRAS} dust masses to estimate the current molecular gas contents, assuming a G/D of $1190 \pm 410$. The median molecular gas fraction (i.e. molecular gas mass divided by stellar mass) is $\sim$1.3 per cent. The error bars incorporate both the uncertainty on the G/D value and the errors in the \emph{IRAS}-derived dust masses. It is worth noting that this value is well within the distribution of molecular gas fractions found by \citet{you11} in a volume-limited sample of \emph{normal} ETGs in the ATLAS$^{\textnormal{3D}}$ survey of very nearby ETGs.

Since star formation is ongoing in these galaxies, we expect the current molecular gas masses calculated above to be lower than the initial molecular gas contents. Given a characteristic exponential time-scale ($\tau_d$) for the gas depletion (which is also the star formation time-scale), the current molecular gas fraction ($g_0$) is related to the initial value ($g_i$) by
\begin{eqnarray}
g_0 = g_\textrm{i}e^{(-T/\tau_d)},
\label{equation:depletion}
\end{eqnarray}

\noindent where $T$ is the look-back time to the onset of star formation (i.e. the starburst age). Using equation~(\ref{equation:age}) to estimate $T$, and adopting a value of 0.1 Gyr for the characteristic time-scale of the recent star formation episodes in ETGs \citep[see][]{sch07}, we estimate $g_i$ in each of our D-ETGs. These initial molecular gas fractions are presented in the bottom panel of Fig.~\ref{fig:gf}. The estimated initial gas fractions are small, with a median value of around 4 per cent. Recent work \citep[see e.g.][]{kav09,kav11} has suggested that the merger activity (and associated star formation) in nearby ETGs largely involves minor mergers between dry ETGs and gas-rich satellites with mass ratios between 1:4 and 1:10. If the \emph{IRAS}-detected D-ETGs indeed formed via this channel, then a typical initial gas fraction of 4 per cent implies that the gas fractions of the accreted satellites are likely to be between $\sim$20 and $\sim$44 per cent (assuming that the ETGs themselves are gas free). It is worth noting that these empirical results are consistent with recent numerical work which suggests that the satellites fuelling the star formation in nearby ETGs should have molecular gas fractions greater than 20 per cent \citep{kav09}. It should also be noted that the molecular gas analysis is only representative of the 15 per cent of D-ETGs that are detected by \emph{IRAS}. A more robust statistical analysis requires infrared data on the entirety of the D-ETG sample studied here.


\section{Summary}
\label{section:summary}
We have explored the properties of dust and associated molecular gas in ETGs, by studying 352 nearby ($0.01<z<0.07$) ETGs with prominent dust features in the SDSS DR7. The galaxies were identified through direct visual inspection of SDSS images using the GZ2 project. Our analysis has focused on comparing the local environment, star formation and nuclear activity of these D-ETGs to those of a control sample drawn from the general ETG population. We have then explored the properties and origin of the dust and associated molecular gas in these systems, and studied the processes that lead to their formation.

Two-thirds of the D-ETGs show morphological disturbances at the depth of the standard SDSS images, suggesting a merger origin for these galaxies. D-ETGs reside preferentially in low-density environments, with 80 per cent inhabiting the field (compared to 60 per cent of their control-sample counterparts) and less than 2 per cent inhabiting clusters (compared to 10 per cent of their control-sample counterparts). D-ETGs show bluer UV-optical colours compared to the control ETG population, indicating the presence of enhanced levels of star formation activity. An analysis using optical emission-line ratios indicates that the fraction of objects classified as star forming and composite are, respectively, factors of 3 and 8 higher in the D-ETGs compared to their control-sample counterparts. The fraction of Seyferts is more than an order of magnitude higher, while the fraction of galaxies classified as quiescent is correspondingly an order of magnitude smaller in the D-ETG population compared to that in the control sample. The presence of prominent dust features in an ETG therefore significantly increases its chances of hosting both star formation and nuclear activity, with nuclear activity being the dominant gas ionization mechanism.

We have explored the dust content of our D-ETGs, both in terms of the clumpy dust mass contained in the large-scale dust features (visible in the optical SDSS images) and in terms of the diffuse dust that may exist in the ISM. The clumpy dust masses span the range $10^{4.5}$--$10^{6.5}$M$_{\odot}$ and are typically several factors higher than the maximum dust contribution expected from stellar mass loss from the old, underlying stellar populations that dominate ETGs. The clumpy dust mass does not correlate with the blue luminosity of the galaxies, indicating that galactic-scale cooling flows are unlikely to be responsible for the dust content. Similarly, no correlation is found between the clumpy dust content and the age of the recent starburst, suggesting that the dust is not produced in situ from the star formation triggered by the merger event, but is rather accreted directly during the merger.

The clumpy dust masses are strictly lower limits to the total dust masses because they do not include diffuse dust which may be intermixed with the stars and is essentially invisible in the optical images. In the 15 per cent of our D-ETGs that are detected by \emph{IRAS}, we have estimated the `total' (clumpy + diffuse) dust mass using the \emph{IRAS} 60- and 100-$\mu$m fluxes. The average discrepancy between the clumpy and \emph{IRAS}-derived dust mass in these D-ETGs is a factor of 4.5. In other words, only $\sim$20 per cent of the dust mass in these objects is contained in the large-scale dust features that are observed in the optical images. Nevertheless, the \emph{IRAS} dust masses show the same lack of correlation with galaxy properties and starburst ages as their clumpy counterparts.

The correspondence between dust and cold gas allows us to explore the properties of the cold gas reservoirs in our D-ETGs. Restricting ourselves to the \emph{IRAS}-detected galaxies (in which we have a reasonable estimates of the total dust mass), we employ the average molecular G/D of ETGs in the literature ($\sim 1190$) to estimate the current molecular gas fractions of these objects. The median value of the current gas fraction is $\sim$ 1.3 per cent. Given the current molecular gas fraction, an estimate of the age of the starburst and a characteristic gas depletion (star formation) time-scale, we have estimated the \emph{initial} molecular gas fraction in each \emph{IRAS}-detected D-ETG, with a median value of $\sim$4 per cent. Given recent work that suggests that the merger activity in nearby ETGs largely involves minor mergers with mass ratios between 1:10 and 1:4 \citep[e.g.][]{kav09}, a typical initial molecular gas fraction of 4 per cent suggests that the gas-rich satellites accreted by the \emph{IRAS}-detected ETGs are likely to have had molecular gas fractions between $\sim$20 and 44 per cent (assuming that the ETGs themselves are gas and dust free).

While recent UV-optical studies have measured the continuous (minor-merger-driven) star formation in ETGs over the last 8 billion years, less is known about the dust and associated gas that drives this star formation. This paper has offered insights into the ISM of nearby ETGs that are merger remnants and, in particular, the initial gas masses plausibly brought in by the satellites that drive the star formation. The galaxy sample studied here provides an ideal basis for future studies of D-ETGs using new instruments such as \emph{Herschel} and the Atacama Large Millimeter Array (ALMA) to probe the properties of the ISM of nearby ETGs and achieve a fuller understanding of the gas that drives the star formation in these galaxies at late epochs.


\section*{Acknowledgements}
We are grateful to the anonymous referee for many insightful comments, that improved the quality of the original manuscript. SK acknowledges fellowships from the Royal Commission for the Exhibition of 1851, Imperial College London, Worcester College, Oxford and support from the BIPAC institute at Oxford. MB acknowledges support through STFC rolling grant PP/E001114/1. SS thanks the Australian Research Council and New College, Oxford for research fellowships.
 



\begin{thebibliography}{99}
\bibitem[\protect\citeauthoryear{Abazajian et al.}{2009}]{aba09} Abazajian K.~N. et al., 2009, \apjs, 182, 543
\bibitem[\protect\citeauthoryear{Annibali et al.}{2010}]{ann10} Annibali F., Bressan A., Rampazzo R., Zeilinger W.~W., Vega O., Panuzzo P., 2010, \aap, 519, A40
\bibitem[\protect\citeauthoryear{Baldwin, Phillips \& Terlevich}{Baldwin et al.}{1981}]{bal81} Baldwin J.~A., Phillips M.~M., Terlevich R., 1981, \pasp, 93, 5
\bibitem[\protect\citeauthoryear{Bell et al.}{2003}]{bel03} Bell E. F., McIntosh D.~H., Katz N., Weinberg M.~D., 2003, \apjs, 149, 289
\bibitem[\protect\citeauthoryear{Bell et al.}{2004}]{bel04} Bell E. F. et al., 2004, \apj, 608, 752
\bibitem[\protect\citeauthoryear{Bernardi et al.}{2003}]{ber03} Bernardi M. et al., 2003, \aj, 125, 1882
\bibitem[\protect\citeauthoryear{Bertola}{1987}]{ber87} Bertola F., 1987, in de Zeeuw P. T., ed., Proc IAU Symp. 127, Structure and Dynamics of Elliptical Galaxies. Reidel, Dordrecht, p. 135
\bibitem[\protect\citeauthoryear{Binney \& Tremaine}{1987}]{bin87} Binney J., Tremaine S., 1987, Galactic Dynamics. Princeton Univ. Press, Princeton
\bibitem[\protect\citeauthoryear{Blanton \& Roweis}{2007}]{bla07} Blanton M.~R., Roweis S., 2007, \aj, 133, 734
\bibitem[\protect\citeauthoryear{Blanton et al.}{2003}]{bla03} Blanton M.~R. et al., 2003, \aj, 125, 2348
\bibitem[\protect\citeauthoryear{Bower, Lucey \& Ellis}{Bower et al.}{1992}]{bow92} Bower R.~G., Lucey J.~R., Ellis R., 1992, \mnras, 254, 589
\bibitem[\protect\citeauthoryear{Bregman, Hogg \& Roberts}{Bregman et al.}{1992}]{bre92} Bregman J.~N., Hogg D.~E., Roberts M.~S., 1992, \apj, 387, 484
\bibitem[\protect\citeauthoryear{Calura, Pipino \& Matteucci}{Calura et al.}{2008}]{cal08} Calura F., Pipino A., Matteucci F., 2008, \aap, 479, 669
\bibitem[\protect\citeauthoryear{Canizares, Fabbiano \& Trinchieri}{Canizares et al.}{1987}]{can87} Canizares C.~R., Fabbiano G., Trinchieri G., 1987, \apj, 312, 503
\bibitem[\protect\citeauthoryear{Combes, Young \& Bureau}{Combes et al.}{2007}]{com07} Combes F., Young L.~M., Bureau M., 2007, \mnras, 377, 1795
\bibitem[\protect\citeauthoryear{Crockett et al.}{2011}]{cro11} Crockett R.~M. et al., 2011, \apj, 727, 115
\bibitem[\protect\citeauthoryear{Davis et al.}{2011}]{dav11} Davis T.~A. et al., 2011, \mnras, 417, 882
\bibitem[\protect\citeauthoryear{Draine \& Lee}{1984}]{dra84} Draine B.~T., Lee H.~M., 1984, \apj, 285, 89
\bibitem[\protect\citeauthoryear{Draine \& Salpeter}{1979a}]{dra79a} Draine B.~T., Salpeter E.~E., 1979a, \apj, 231, 77
\bibitem[\protect\citeauthoryear{Draine \& Salpeter}{1979b}]{dra79b} Draine B.~T., Salpeter E.~E., 1979b, \apj, 231, 438
\bibitem[\protect\citeauthoryear{Dwek}{1986}]{dwe86} Dwek E., 1986, \apj, 302, 363
\bibitem[\protect\citeauthoryear{Dwek, Rephaeli \& Mather}{Dwek et al.}{1990}]{dwe90} Dwek E., Rephaeli Y., Mather J.~C., 1990, \apj, 350, 104
\bibitem[\protect\citeauthoryear{Ebneter, Davis \& Djorgovski}{Ebneter et al.}{1988}]{ebn88} Ebneter K., Davis M., Djorgovski S., 1988, \aj, 95, 422
\bibitem[\protect\citeauthoryear{Faber \& Gallagher}{1976}]{fab76} Faber S.~M., Gallagher J.~S., 1976, \apj, 204, 365
\bibitem[\protect\citeauthoryear{Faber et al.}{1997}]{fab97} Faber S.~M. et al., 1997, \aj, 114, 1771
\bibitem[\protect\citeauthoryear{Faber et al.}{2007}]{fab07} Faber S.~M. et al., 2007, \apj, 665, 265
\bibitem[\protect\citeauthoryear{Ferreras et al.}{2009}]{fer09} Ferreras I., Lisker T., Pasquali A., Khochfar S., Kaviraj S., 2009, \mnras, 396, 1573
\bibitem[\protect\citeauthoryear{Finkelman et al.}{2008}]{fin08} Finkelman I. et al., 2008, \mnras, 390, 969
\bibitem[\protect\citeauthoryear{Forbes}{1991}]{for91} Forbes D.~A., 1991, \mnras, 249, 779
\bibitem[\protect\citeauthoryear{Forbes, Ponman \& Brown}{Forbes et al.}{1998}]{for98} Forbes D.~A., Ponman T.~J., Brown R.~J.~N., 1998, \apj, 508, L43
\bibitem[\protect\citeauthoryear{Fukugita et al.}{2004}]{fuk04} Fukugita M., Nakamura O., Turner E.~L., Helmboldt J., Nichol R.~C., 2004, \apj, 601, L127
\bibitem[\protect\citeauthoryear{Gordon}{1991}]{gor91} Gordon M.~A., 1991, \apj, 371, 563
\bibitem[\protect\citeauthoryear{Goudfrooij \& de Jong}{1995}]{gou95} Goudfrooij P., de Jong T., 1995, \aap, 298, 784
\bibitem[\protect\citeauthoryear{Goudfrooij et al.}{1994}]{gou94} Goudfrooij P., de Jong T., Hansen L., Norgaard-Nielsen H.~U., 1994, \mnras, 271, 833
\bibitem[\protect\citeauthoryear{Hawarden et al.}{1981}]{haw81} Hawarden T.~G., Longmore A.~J., Tritton S.~B., Elson R.~A.~W., Corwin H.~G., Jr, 1981, \mnras, 196, 747
\bibitem[\protect\citeauthoryear{Henkel \& Wiklind}{1997}]{hen97} Henkel C., Wiklind T., 1997, \ssr, 81, 1
\bibitem[\protect\citeauthoryear{Hildebrand}{1983}]{hil83} Hildebrand R.~H., 1983, \qjras, 24, 267
\bibitem[\protect\citeauthoryear{Jeong et al.}{2007}]{jeo07} Jeong H., Bureau M., Yi S.~K., Krajnovi{\'c} D., Davies R. L., 2007, \mnras, 376, 1021
\bibitem[\protect\citeauthoryear{Jeong et al.}{2009}]{jeo09} Jeong H. et al., 2009, \mnras, 398, 2028
\bibitem[\protect\citeauthoryear{Jogee et al.}{2009}]{jog99} Jogee S. et al., 2009, \apj, 697, 1971
\bibitem[\protect\citeauthoryear{Jorgensen, Franx \& Kjaergaard}{Jorgensen et al.}{1996}]{jor96} Jorgensen I., Franx M., Kjaergaard P., 1996, \mnras, 280, 167
\bibitem[\protect\citeauthoryear{Kauffmann et al.}{2003}]{kau03} Kauffmann G. et al., 2003, \mnras, 346, 1055
\bibitem[\protect\citeauthoryear{Kaviraj}{2010}]{kav10} Kaviraj S., 2010, \mnras, 408, 170
\bibitem[\protect\citeauthoryear{Kaviraj et al.}{2007}]{kav07} Kaviraj S. et al., 2007, \apjs, 173, 619
\bibitem[\protect\citeauthoryear{Kaviraj et al.}{2008}]{kav08} Kaviraj S. et al., 2008, \mnras, 388, 67
\bibitem[\protect\citeauthoryear{Kaviraj et al.}{2009}]{kav09} Kaviraj S., Peirani S., Khochfar S., Silk J., Kay S., 2009, \mnras, 394, 1713
\bibitem[\protect\citeauthoryear{Kaviraj et al.}{2011}]{kav11} Kaviraj S., Tan K.~M., Ellis R.~S., Silk J., 2011, \mnras, 411, 2148
\bibitem[\protect\citeauthoryear{Kewley et al.}{2006}]{kew06} Kewley L. J., Groves B., Kauffmann G., Heckman T., 2006, \mnras, 372, 961
\bibitem[\protect\citeauthoryear{Knapp}{1999}]{kna99} Knapp G.~R., 1999, in Carral P., Cepa J., eds, ASP Conf. Ser. Vol. 163, Star Formation in Early Type Galaxies. Astron. Soc. Pac., San Francisco, p.~119
\bibitem[\protect\citeauthoryear{Knapp \& Rupen}{1996}]{kna96} Knapp G.~R., Rupen M.~P., 1996, \apj, 460, 271
\bibitem[\protect\citeauthoryear{Knapp et al.}{1989}]{kna89} Knapp G.~R., Guhathakurta P., Kim D., Jura M.~A., 1989, \apjs, 70, 329
\bibitem[\protect\citeauthoryear{Knapp, Gunn \& Wynn-Williams}{Knapp et al.}{1992}]{kna92} Knapp G.~R., Gunn J.~E., Wynn-Williams C.~G., 1992, \apj, 399, 76
\bibitem[\protect\citeauthoryear{Krumholz \& McKee}{2005}]{kru05} Krumholz M.~R., McKee C.~F., 2005, \apj, 630, 250
\bibitem[\protect\citeauthoryear{Kwan \& Xie}{1992}]{kwa92} Kwan J., Xie S., 1992, \apj, 398, 105
\bibitem[\protect\citeauthoryear{Lees et al.}{1991}]{lee91} Lees J.~F., Knapp G.~R., Rupen M.~P., Phillips T.~G., 1991, \apj, 379, 177
\bibitem[\protect\citeauthoryear{Leroy et al.}{2008}]{ler08} Leroy A.~K. et al., 2008, \aj, 136, 2782
\bibitem[\protect\citeauthoryear{Lintott et al.}{2008}]{lin08} Lintott C. et al., 2008, \mnras, 389, 1179
\bibitem[\protect\citeauthoryear{Lintott et al.}{2011}]{lin10} Lintott C. et al., 2011, \mnras, 410, 166
\bibitem[\protect\citeauthoryear{L\'opez-Sanjuan et al.}{2011}]{lop10} L\'opez-Sanjuan C., Balcells M., P\'erez-Gonz\'alez P.~G., Barro G., Gallego J., Zamorano J., 2011, \aap, 530, A20
\bibitem[\protect\citeauthoryear{Martin et al.}{2005}]{mar05} Martin D.~C. et al., 2005, \apj, 619, L1
\bibitem[\protect\citeauthoryear{Martin, O'Connell \& Hibbard}{2009}]{mar09} Martin J.~R., O'Connell R.~W., Hibbard J.~E., 2009, in Ch\'avez Dagostino M., Bertone E., Rosa Gonzalez D., Rodgriguez-Merino L.~H., eds, New Quests in Stellar Astrophysics. II. Ultraviolet Properties of Evolved Stellar Populations, Springer, Berlin, p.~83
\bibitem[\protect\citeauthoryear{Mathew \& Brighenti}{2003}]{mat03} Mathews W.~G., Brighenti F., 2003, \araa, 41, 191
\bibitem[\protect\citeauthoryear{Merluzzi}{1998}]{mer98} Merluzzi P., 1998, \aap, 338, 807
\bibitem[\protect\citeauthoryear{Morrissey et al.}{2007}]{mor07} Morrissey P. et al., 2007, \apjs, 173, 682
\bibitem[\protect\citeauthoryear{Mulchaey \& Jeltema}{2010}]{mul10} Mulchaey J.~S., Jeltema T.~E., 2010, \apj, 715, L1
\bibitem[\protect\citeauthoryear{Peirani et al.}{2010}]{pei10} Peirani S., Crockett R.~M., Geen S., Khochfar S., Kaviraj S., Silk J., 2010, \mnras, 405, 2327
\bibitem[\protect\citeauthoryear{Rodgers et al.}{2006}]{rod06} Rodgers C.~T., Cantema R., Smith J.~A., Pierce M.~J., Tucker D.~L., 2006, \aj, 132, 989
\bibitem[\protect\citeauthoryear{Rowan-Robinson}{1986}]{row86} Rowan-Robinson M., 1986, \mnras, 219, 737
\bibitem[\protect\citeauthoryear{Sadler \& Gerhard}{1985}]{sad85} Sadler E.~M., Gerhard O.~E., 1985, \mnras, 214, 177
\bibitem[\protect\citeauthoryear{Sage \& Galletta}{1993}]{sag93} Sage L.~J., Galletta G., 1993, \apj, 419, 544
\bibitem[\protect\citeauthoryear{Saglia et al.}{1997}]{sag97} Saglia R.~P., Colless M., Baggley G. et al., 1997, in Arnaboldi M., Da Costa G.~S., Saha P., eds, ASP Conf. Ser. Vol. 116, The Nature of Elliptical Galaxies. Astron. Soc. Pac., San Francisco, p.~180
\bibitem[\protect\citeauthoryear{Sahu, Pandey \& Kembhavi}{Sahu et al.}{1998}]{sah98} Sahu D.~K., Pandey S.~K., Kembhavi A., 1998, \aap, 333, 803
\bibitem[\protect\citeauthoryear{Salim \& Rich}{2010}]{sal10} Salim S., Rich R.~M., 2010, \apj, 714, L290
\bibitem[\protect\citeauthoryear{Sarzi et al.}{2006}]{sar06} Sarzi M. et al., 2006, \mnras, 366, 1151
\bibitem[\protect\citeauthoryear{Sarzi et al.}{2010}]{sar10} Sarzi M. et al., 2010, \mnras, 402, 2187
\bibitem[\protect\citeauthoryear{Schawinski et al.}{2007}]{sch07} Schawinski K., Thomas D., Sarzi M., Maraston C., Kaviraj S., Joo S., Yi S.~K., Silk J., 2007, \mnras, 382, 1415
\bibitem[\protect\citeauthoryear{Schaye}{2004}]{sch04} Schaye J., 2004, \apj, 609, 667
\bibitem[\protect\citeauthoryear{Schwartz}{1982}]{sch82} Schwartz P.~R., 1982, \apj, 252, 589
\bibitem[\protect\citeauthoryear{Schweizer \& Seitzer}{1992}]{sch92} Schweizer F., Seitzer P., 1992, \aj, 104, 1039
\bibitem[\protect\citeauthoryear{Schweizer et al.}{1990}]{sch90} Schweizer F., Seitzer P., Faber S.~M., Burstein D., Dalle Ore C.~M., Gonzalez J.~J., 1990, \apj, 364, L33
\bibitem[\protect\citeauthoryear{Spitzer}{1978}]{spi78} Spitzer L., 1978, Physical Processes in the Interstellar Medium. Wiley-Interscience, New York
\bibitem[\protect\citeauthoryear{Stewart}{2008}]{ste08} Stewart K.~R., Bullock J.~S., Wechsler R.~H., Maller A.~H., Zentner A.~R., 2008, \apj, 683, 597
\bibitem[\protect\citeauthoryear{Stoughton et al.}{2002}]{sto02} Stoughton C. et al., 2002, \aj, 123, 485
\bibitem[\protect\citeauthoryear{Suh et al.}{2010}]{suh10} Suh H., Jeong H., Oh K., Yi S.~K., Ferreras I., Schawinski K., 2010, \apjs, 187, 374
\bibitem[\protect\citeauthoryear{Thomas, Greggio \& Bender}{Thomas et al.}{1999}]{tho99} Thomas D., Greggio L, Bender R., 1999, \mnras, 302, 537
\bibitem[\protect\citeauthoryear{Thronson et al.}{1989}]{thr89} Thronson H.~A., Jr., Tacconi L., Kenney J.~D.~P., Greenhouse M.~A., Margulis M., Tacconi-Garman L., Young J.~S., 1989, \apj, 344, 747
\bibitem[\protect\citeauthoryear{Tomita et al.}{2000}]{tom00} Tomita A., Aoki K., Watanabe M., Takata T., Ichikawa S.-i., 2000, \aj, 120, 123
\bibitem[\protect\citeauthoryear{Trager et al.}{2000a}]{tra00a} Trager S.~C., Faber S.~M., Worthey G., Gonz\'alez J.~J., 2000a, \aj, 119, 1645
\bibitem[\protect\citeauthoryear{Trager et al.}{2000b}]{tra00b} Trager S.~C., Faber S.~M., Worthey G., Gonz\'alez J.~J., 2000b, \aj, 120, 165
\bibitem[\protect\citeauthoryear{Tran et al.}{2001}]{tra01} Tran H.~D., Tsvetanov Z., Ford H.~C., Davies J., Jaffe W., van den Bosch F.~C., Rest A., 2001, \aj, 121, 2928
\bibitem[\protect\citeauthoryear{Tubbs}{1980}]{tub80} Tubbs A.~D., 1980, \apj, 241, 969
\bibitem[\protect\citeauthoryear{van Dokkum \& Franx}{1995}]{van95} van Dokkum P.~G., Franx M., 1995, \aj, 110, 2027
\bibitem[\protect\citeauthoryear{van Dokkum \& Franx}{1996}]{van96} van Dokkum P.~G., Franx M., 1996, \mnras, 281, 985
\bibitem[\protect\citeauthoryear{Vanden Berk et al.}{2001}]{van01} Vanden Berk D.~E. et al., 2001, \aj, 122, 549
\bibitem[\protect\citeauthoryear{Veilleux \& Osterbrock}{1987}]{vei87} Veilleux S., Osterbrock D.~E., 1987, \apjs, 63, 295
\bibitem[\protect\citeauthoryear{Wang \& Rowan-Robinson}{2009}]{wan09} Wang L., Rowan-Robison M., 2009, \mnras, 398, 109
\bibitem[\protect\citeauthoryear{Wang, Kenney \& Ishizuki}{Wang et al.}{1992}]{wan92} Wang Z., Kenney J.~D.~P., Ishizuki S., 1992, \aj, 104, 2097
\bibitem[\protect\citeauthoryear{Whitaker \& van Dokkum}{2008}]{whi08} Whitaker K.~E., van Dokkum P.~G., 2008, \apj, 676, L105
\bibitem[\protect\citeauthoryear{Wiklind \& Henkel}{1989}]{wik89} Wiklind T., Henkel C., 1989, \aap, 225, 1
\bibitem[\protect\citeauthoryear{Wiklind \& Henkel}{1995}]{wik95} Wiklind T., Henkel C., 1995, \aap, 297, L71
\bibitem[\protect\citeauthoryear{Wong \& Blitz}{2002}]{won02} Wong T., Blitz L, 2002, \apj, 569, 157
\bibitem[\protect\citeauthoryear{Yang et al.}{2007}]{yan07} Yang X., Mo H.~J., van den Bosch F.~C., Pasquali A., Li C., Barden M., 2007, \apj, 671, 153
\bibitem[\protect\citeauthoryear{Yang, Mo, \& van den Bosch}{Yang et al.}{2008}]{yan08} Yang X., Mo H.~J., van den Bosch F.~C., 2008, \apj, 676, 248
\bibitem[\protect\citeauthoryear{Yi et al.}{2005}]{yi05} Yi S.~K. et al., 2005, \apj, 619, L111
\bibitem[\protect\citeauthoryear{York et al.}{2000}]{yor00} York D.~G. et al. (SDSS Collaboration) 2000, \aj, 120, 1579
\bibitem[\protect\citeauthoryear{Young et al.}{1989}]{you89} Young J.~S., Xie S., Kenney J.~D.~P., Rice W.~L., 1989, \apjs, 70, 699
\bibitem[\protect\citeauthoryear{Yong et al.}{2011}]{you11} Young L.~M. et al., 2011, \mnras, 414, 940
\bibitem[\protect\citeauthoryear{Zeilinger, Bertola \& Galletta}{Zeilinger et al.}{1990}]{zei90} Zeilinger W.~W., Bertola F., Galletta G., 1990, in Bussoletti E., Vittone A.~A., eds, Astrophys. Space Sci. Libr. Vol. 165, Dusty Objects in the Universe. Kluwer, Dordrecht, p.~227
\end{thebibliography}
\end{document}